\documentclass[showpacs,amsmath,superscriptaddress,reprint,aps]{revtex4-1}

\usepackage[dvipdfmx]{graphicx}
\usepackage{hyperref}
\usepackage{color}
\usepackage{float}

\begin{document}

\title
{Deterministic transfer of optical-quality carbon nanotubes for atomically defined technology}
\author{K.~Otsuka}
\affiliation{Nanoscale Quantum Photonics Laboratory, RIKEN Cluster for Pioneering Research, Saitama 351-0198, Japan}
\author{N.~Fang}
\affiliation{Nanoscale Quantum Photonics Laboratory, RIKEN Cluster for Pioneering Research, Saitama 351-0198, Japan}
\author{D.~Yamashita}
\affiliation{Quantum Optoelectronics Research Team, RIKEN Center for Advanced Photonics, Saitama 351-0198, Japan}
\author{T.~Taniguchi}
\affiliation{International Center for Materials Nanoarchitectonics, National Institute for Materials Science, Ibaraki 305-0044, Japan}
\author{K.~Watanabe}
\affiliation{Research Center for Functional Materials, National Institute for Materials Science, Ibaraki 305-0044, Japan}
\author{Y.~K.~Kato}
\email[Corresponding author. ]{yuichiro.kato@riken.jp}
\affiliation{Nanoscale Quantum Photonics Laboratory, RIKEN Cluster for Pioneering Research, Saitama 351-0198, Japan}\affiliation{Quantum Optoelectronics Research Team, RIKEN Center for Advanced Photonics, Saitama 351-0198, Japan}

\begin{abstract}
When continued device scaling reaches the ultimate limit imposed by atoms, technology based on atomically precise structures is expected to emerge. Device fabrication will then require building blocks with identified atomic arrangements and assembly of the components without contamination. Here we report on a versatile dry transfer technique for deterministic placement of optical-quality carbon nanotubes. Single-crystalline anthracene is used as a medium which readily sublimes by mild heating, leaving behind clean nanotubes and thus enabling bright photoluminescence. We are able to position nanotubes of a desired chirality with a sub-micron accuracy under in-situ optical monitoring, thereby demonstrating deterministic coupling of a nanotube to a photonic crystal nanobeam cavity. A cross junction structure is also designed and constructed by repeating the nanotube transfer, where intertube exciton transfer is observed. Our results represent an important step towards development of devices consisting of atomically precise components and interfaces.
\end{abstract}

\maketitle
\section{Introduction}
``There's plenty room at the bottom.'' Since the first notion of nanotechnology~\cite{Feynman:1960}, device miniaturization has been the driving force for technological evolution.  Rapid progress of nanofabrication techniques has pushed forward the Moore's Law, exponentially increasing the computation powers of semiconductor microprocessors that now form the foundations for large-scale simulations and artificial intelligence. Fueled by scientific curiosity and technological thirst, efforts for further scaling will undoubtedly continue until we reach the ultimate limit where devices are constructed of components and interfaces with atomic precision: Atomically defined technology.

Recent developments in heterostructures of two-dimensional (2D) materials~\cite{Novoselov:2004,Novoselov:2005, Dean:2010} have highlighted novel exotic phenomena arising from the intricacies of precise atomic arrangements. Twisted bilayer graphene exhibits superconductivity at the magic angle~\cite{Cao:2018}, whereas twisted homobilayer of 2D semiconductor with a long-period moir\'e superlattice gives rise to incompressible Mott-like states of electrons~\cite{Shimazaki:2020}. In addition, high-temperature condensation of interlayer excitons has been observed in aligned MoSe$_2$-WSe$_2$ atomic double layers~\cite{Wang:2019}, where control over the interlayer spacing by atomically thin insulators also plays a key role.

To harness the full functionality and performance of atomically precise structures, systems with more complexity than stacked layers would be necessary. It is, however, not an easy task to assemble atomically defined building blocks for real world applications. Organic molecules allow for tailoring their structures at the atomic level~\cite{Park:2000,Perrin:2015}, but manipulation of single molecules requires low temperature and ultrahigh vacuum~\cite{Moresco:2001}. Although it is possible to integrate individual semiconductor nanostructures into electronic and photonic devices~\cite{Klein:1997,Santori:2002}, it still remains a challenge to reproducibly prepare identical structures with atomic precision. 

Carbon nanotubes (CNTs) offer a unique position in this context. Their atomic arrangements can be specified by chirality which is a combination of two integers defining the geometry of the roll-up vector, and they can be individually addressed to construct various nanoscale devices such as single-electron transistors ~\cite{Postma:2001} and light-emitting diodes~\cite{Mueller:2010}. Simultaneous control over chirality, position, and orientation of CNTs would enable fabrication of devices utilizing atomically precise components, given that their intrinsic properties are preserved during assembly. Picking up air-suspended tubes and re-suspending them is an ideal method for keeping the surface pristine~\cite{Huang:2005, Abrams:2007, Wu:2010, Waissman:2013}, but the necessity for three-dimensional structures imposes limitations on device construction. In addition, state-of-the-art growth control over tube diameter and length on substrate surfaces~\cite{Kocabas:2005, Yang:2014, Zhang:2017nature} cannot be utilized. A different approach for manipulating chirality-identified tubes with negligible contamination would be required to achieve the flexibility needed for integrating multiple elements on demand.

Here we demonstrate deterministic transfer of optical-quality CNTs by utilizing anthracene as a sacrificial material. The use of large anthracene crystals allows transfer onto arbitrary substrates, and removal of anthracene through sublimation recovers the CNT luminescence intensity to the level of as-grown tubes. By monitoring the PL of CNTs during the transfer, we can select nanotubes of a desired chirality and position them with sub-micron spatial accuracy. As a demonstration of the chirality-on-demand transfer technique, we perform deterministic coupling to a photonic crystal nanobeam cavity. Furthermore, we design and assemble a cross junction from semiconducting CNTs to observe intertube exciton transport.

\section{Results and discussion}
\paragraph*{Anthracene-assisted nanotube transfer.}

\begin{figure*}
\includegraphics{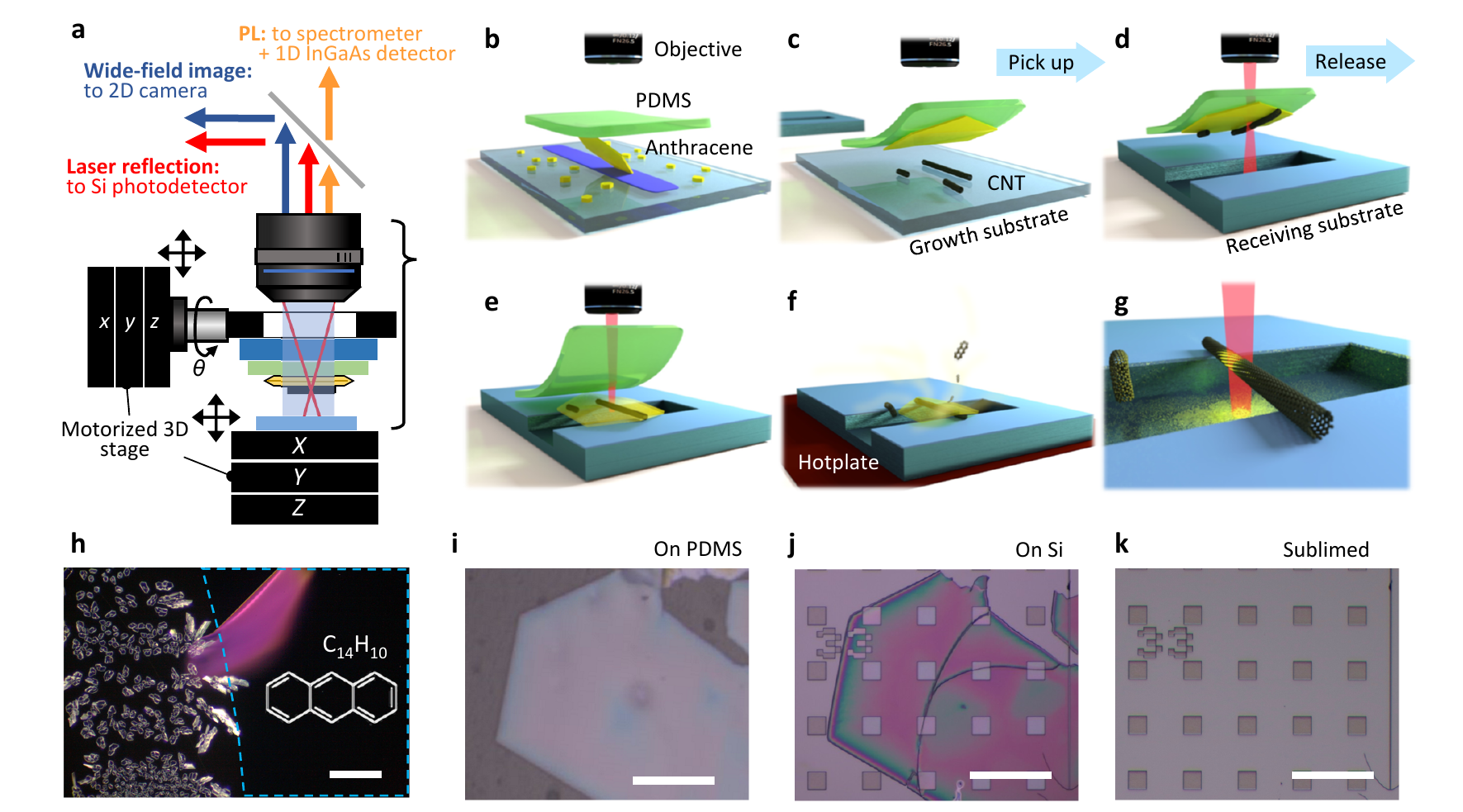}\
\caption{
\label{Fig1} Procedure for anthracene-assisted dry transfer of carbon nanotubes. (a) Schematic of the transfer system that simultaneously monitors the nanotube PL, the laser reflectivity, and the wide-field image. (b) Anthracene crystals are grown on glass slides through in-air sublimation and are picked up with a PDMS stamp. (c) CNTs are picked up with the PDMS/anthracene stamp, (d) followed by a PL measurement to locate CNTs of interest. (e) Anthracene/CNTs are released on a receiving substrate by peeling off the stamp. (f) In-air heating of the substrate removes the anthracene crystals and (g) leaves behind the CNTs alone. (h--k) Typical optical images of anthracene crystals grown on a glass slide (h), picked up on a PDMS (i), transferred to a Si chip with an array of pits (j), and after sublimation in air (k). Region outlined by a blue dashed box in (h) is coated with marker ink. Scale bars are 100~$\mu$m.
}
\end{figure*}

Figure~\ref{Fig1} illustrates the steps of the anthracene-assisted transfer of individual CNTs. To pick up long CNTs and release them on arbitrary substrates with minimum contamination, we use thin and large-area organic crystals that do not require wet processes or high-temperature treatments. Anthracene, which consists of three benzene rings, meets such requirements~\cite{Yulaev:2016, Chen:2017}. We expect the similarity of molecular structure between anthracene and CNTs to lead to strong $\pi$-$\pi$ interactions~\cite{Chen:2001}, possibly improving the efficiency of picking up CNTs from the growth substrates. The charge-neutral nature of anthracene~\cite{Lu:2006} should suppress exciton quenching, and we anticipate PL of CNTs to remain bright.

As shown in Figure~\ref{Fig1}(a), we have built a transfer system where the nanotube PL, the laser reflection, and the wide-field optical image are monitored during the nanotube transfer process. An anthracene single crystal is picked up with a glass-supported polydimethylsiloxane (PDMS) sheet (Gelfilm by Gelpak\textregistered)~\cite{Sundar:2004} (Figure~\ref{Fig1}(b)). CNTs are picked up by pressing the anthracene/PDMS stamp against a substrate with as-grown CNTs (Figure~\ref{Fig1}(c)), followed by quick separation ($>$10~mm/s) so that the anthracene crystal remains attached to the PDMS sheet~\cite{Meitl:2005}. PL mapping of CNTs on the anthracene crystal is performed when a CNT of a specific chirality needs to be selected (Figure~\ref{Fig1}(d)). The stamp is then pressed on a receiving substrate. By slowly peeling off the PDMS ($<$0.2~$\mu$m/s), the anthracene crystal with the CNTs are released on the substrate (Figure~\ref{Fig1}(e)). Sublimation of anthracene in air leaves behind clean CNTs on any substrate (Figures~\ref{Fig1}(f) and (g)) because contamination from solvents is absent in the all-dry process. Figures~\ref{Fig1}(h), (i), (j), and (k) show the optical microscopy images of anthracene crystals during the transfer steps, corresponding to the schematics in Figures~\ref{Fig1}(b), (c), (e), and (g), respectively. Being a single crystal, the anthracene stamp can be easily handled and transferred without significant damage even over trenches and pits, and can be sublimed after a typical heating process at 110$^\circ$C for 10~min.

\paragraph*{Clean and efficient transfer of CNTs from quartz.}

\begin{figure*}
\includegraphics{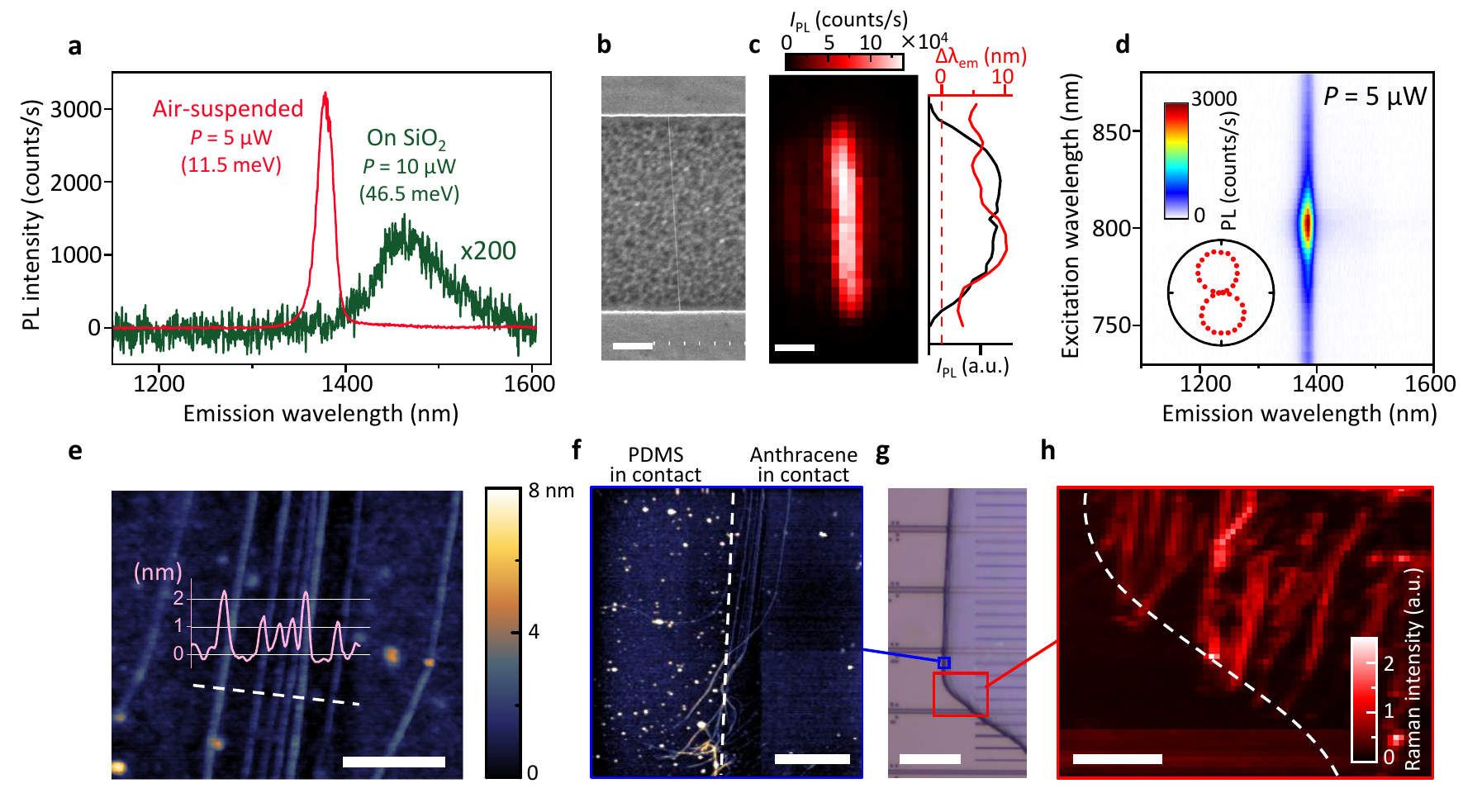}
\caption{
\label{Fig2} Bright PL of a transferred CNT. (a) PL spectra of the CNT transferred from a quartz substrate. Red and green spectra are measured from the same CNT at an air-suspended region and at a region in contact with the SiO$_2$ surface, respectively. Numbers in parentheses represent the full width at half maximum. (b) SEM image of the suspended region and (c) the corresponding PL image at $\lambda_\mathrm{em}=1379$~nm. The peak shift $\Delta \lambda _\mathrm{em}$ (red) from as-grown tubes~\cite{Ishii:2015} and $I_\mathrm{PL}$ (black) along the tube axis are shown in the right panel. (d) PLE spectrum from which chirality is determined to be (9,8). Inset: polarization angle dependence of  $I_\mathrm{PL}$. (e,f) AFM image of the aligned CNTs measured on the SiO$_2$ surface after sublimation of an anthracene crystal. In (f), the left and right sides are in contact with PDMS and anthracene, respectively. (g) Optical image of the anthracene crystal with CNTs from quartz. (h) Raman mapping showing G-band intensity in the region outlined by a red box in (g). Scale bars in (b), (c), (e), (f), (g), and (h) are 1, 1, 0.5, 2, 50, and 10~$\mu$m, respectively.
}
\end{figure*}

We first apply the anthracene-assisted method for CNTs grown on a quartz substrate~\cite{Kocabas:2005}, where the CNTs grow parallel to each other with lengths exceeding 100~$\mu$m. We transfer the CNTs from the quartz substrate to trenches etched on a SiO$_2$/Si substrate. Figure~\ref{Fig2}(a) shows typical PL spectra from a transferred CNT with a total length of $\sim$50~$\mu$m, which are measured at an air-suspended region over a 5-$\mu$m-wide trench (red) and at an on-substrate region (green). The PL efficiency of the suspended region is higher by $\sim$250-fold than the on-substrate region of the same CNT. The absence of SiO$_2$ underneath also narrows and blueshifts the emission spectra by 4-fold and 55~meV, respectively. In Figure~\ref{Fig2}(b), a scanning electron microscopy (SEM) image displays uniformity along the entire length of the air-suspended CNT. In a PL image plotting the integrated PL intensity $I_\mathrm{PL}$ over a $\sim$50~nm spectral window centered at the emission wavelength $\lambda_\mathrm{em}$ (Figure~\ref{Fig2}(c)), we also observe uniform optical response. The intratube variation of the E$_{11}$ peak positions (standard deviation of 2.3~nm) is smaller than the intertube variation in as-grown CNTs with the same chirality (standard deviation of 3.8~nm) reported in a previous study~\cite{Ishii:2015}. We assign the tube chirality to be (9,8) from the PL excitation (PLE) map (Figure~\ref{Fig2}(d)).

Insignificant residue on CNTs and the substrate surface is confirmed in an atomic force microscopy (AFM) image of CNTs, and preservation of CNT alignment after the transfer is observed (Figure~\ref{Fig2}(e)). The measured heights of the CNTs fall within the typical diameter range of those on the growth substrates~\cite{Otsuka:2018}. It is noteworthy that the anthracene stamp gives rise to only one tenth as much contamination as a PDMS stamp (Figure~\ref{Fig2}(f)), suggesting an advantage also for transferring 2D materials.

The use of anthracene as a medium drastically improves the overall transfer yield compared with the transfer using PDMS alone. Figure~\ref{Fig2}(g) shows an optical image of the anthracene crystal with CNTs underneath. G-band Raman mapping in Figure~\ref{Fig2}(h) clearly indicates that CNTs are transferred only through anthracene mediation. The improved transfer yield of the anthracene-assisted method may partially originate from the fact that 100\% of the picked-up CNTs are released on the receiving substrates together with the anthracene crystals. We should note that the pick-up efficiency by the anthracene crystals is much higher for small diameter CNTs ($\sim$1~nm), but the diameter range of CNTs on quartz under the current growth condition ($1.42\pm0.17$~nm~\cite{Otsuka:2018}) lies outside of the optimal range (Figure~S3). Combined with the strong tube-substrate interaction that spontaneously aligns CNTs during the growth, the fraction of CNTs actually picked up is less than 1\%. Optimization of the growth conditions should improve the pickup efficiency and further improve the transfer yield.

\paragraph*{High optical quality of the transferred CNTs.}

\begin{figure}
\includegraphics {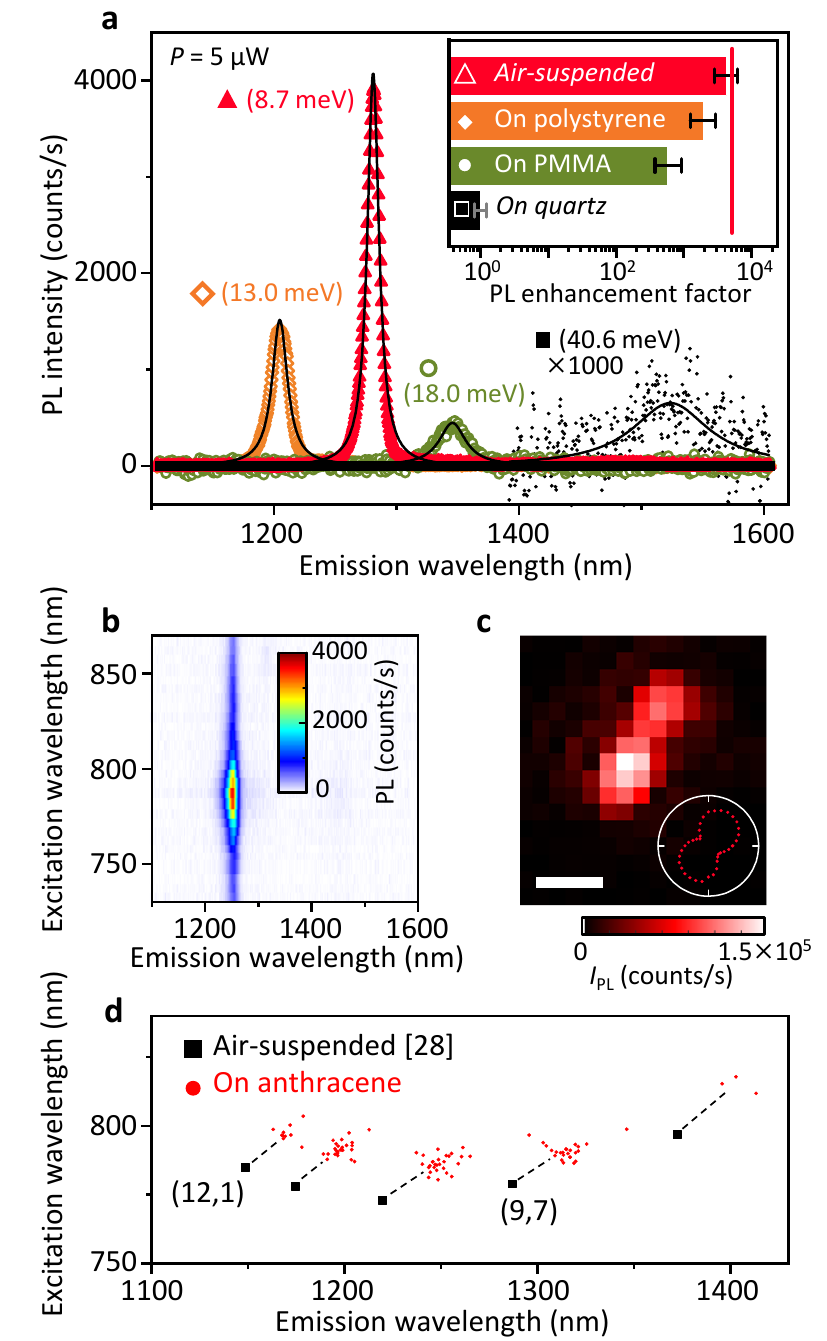}
\caption{
\label{Fig3} PL properties of CNTs on various surfaces. (a) Typical PL spectra with top $\sim$0.1\% of integrated PL intensity $I_\mathrm{PL}$ for each sample at $P=5~\mu$W. Black solid lines are Lorentzian fits. Numbers in parentheses represent the full width at half maximum. Inset: PL enhancement factor in comparison with CNTs on quartz, at $I_\mathrm{PL}$ with a relative frequency of $10^{-3}$. Error bars are the $I_\mathrm{PL}$ ranges that appear twice or half as frequently. Red vertical line represents the transferred air-suspended CNT in Figure~\ref{Fig2}. The sample names in italic represent as-grown CNTs; otherwise, the CNTs are transferred via anthracene. (b) Typical PLE spectrum and (c) PL image at $\lambda_\mathrm{em} =1253~\mathrm{nm}$ of a (10,5) CNT picked up on anthracene crystals. Inset: Polarization dependence of $I_\mathrm{PL}$. Scale bar is 1~$\mu$m. (d) PLE peak positions of CNTs on anthracene (red dots), which are redshifted compared with as-grown suspended CNTs (black squares)~\cite{Ishii:2015}. $P=100~\mu$W is used in (b--d).
}
\end{figure}

The CNTs transferred over trenches should fluoresce as brightly as unprocessed ones in the absence of contamination. To evaluate the PL properties of the CNT shown in Figure~\ref{Fig2}(a), we perform PL measurements on typical substrates used for growth and photonic applications of CNTs; specifically, we use as-grown CNTs on quartz substrates~\cite{Kocabas:2005} or over trenches, and transferred CNTs on poly(methyl methacrylate) (PMMA)~\cite{Legrand:2013} or polystyrene~\cite{He:2017} (see Methods section). We measure $>$10,000 spectra on each sample to perform statistical characterization.

Figure~\ref{Fig3}(a) shows representative PL spectra of the brightest CNTs within each sample. As-grown CNTs on quartz show a broad peak with very small $I_\mathrm{PL}$ due to the strong interaction with the substrate. The PL properties on the polymers are superior to those on quartz and SiO$_2$ (Figure~S6) even in comparison with the previous study~\cite{Schweiger:2015} likely due in part to the absence of wet processes and high-temperature treatments (see Figure~S8). Not surprisingly, the as-grown suspended CNTs fluoresce most brightly and sharply out of all substrates studied.

The influence of each substrate observed in the representative spectra is further confirmed by statistical analysis. The majority of the measured spectra show only noise, implying each peak originates from at most one nanotube (see Figure~S6). We therefore consider $I_\mathrm{PL}$ above the noise floor to be representing the brightness of individual CNTs, whose histogram follows a log-normal distribution in all the samples reflecting the stochastic formation of exciton quenching sites (see Figure~S7). When $I_\mathrm{PL}$ at a relative frequency of $10^{-3}$ is used for comparison, the PL enhancement factor of the CNTs on the polymers with respect to those on quartz ranges between two to three orders of magnitude  (the inset in Figure~\ref{Fig3}(a)). Air-suspended CNTs have $I_\mathrm{PL}$ greater by $>$4,000-fold than unprocessed tubes on quartz. Importantly, the CNT shown in Figure~\ref{Fig2}, whose $I_\mathrm{PL}$ is indicated by the red vertical line in the inset of Figure~\ref{Fig3}(a), even outperforms the best as-grown suspended CNTs. It is somewhat surprising that PL recovery ($\sim$5,000 fold) through the transfer from quartz surface to the trench results in intensity matching as-grown tubes. We note that statistical evaluation for such transferred CNTs requires further improvement of nanotube pick-up efficiency through diameter optimization.

We also find that the bright PL is accompanied by narrow spectral linewidths with small statistical variance (see Figure~S6). The air-suspended CNTs show the sharpest peaks, while CNTs on polymers have linewidths equivalent to those in aqueous agarose gels~\cite{Cambre:2012}.  A slightly broadened linewidth is observed for the air-suspended CNT transferred from quartz in Figure~\ref{Fig2} compared with typical as-grown ones, which may indicate the presence of anthracene residue after the sublimation. The high PL intensity of the transferred CNT comparable to as-grown CNTs suggests that anthracene residue has a negligible effect on PL intensity.

If PL intensity also remains high for nanotubes on anthracene, chirality assignment can be done after picking up the CNTs. As the PL of CNTs on the growth substrate surface is barely detectable, the capability for assigning the tube chirality would be an important factor towards deterministic transfer. We perform PL measurements for the CNTs supported by anthracene crystals over the pits (Figure~\ref{Fig1}(j)). Figure~\ref{Fig3}(b) shows a typical PLE map of a CNT on anthracene, peaking as sharply as the air-suspended ones at the E$_{11}$ and E$_{22}$ resonances. The tube orientation is revealed by both the PL image (Figure~\ref{Fig3}(c)) and the polarization dependence of $I_\mathrm{PL}$ (the inset). As shown in Figure~\ref{Fig3}(d), chiralities of such CNTs can be assigned from PLE spectra prior to releasing onto receiving substrates.

\paragraph*{Deterministic transfer under PL monitoring.}

\begin{figure*}
\includegraphics{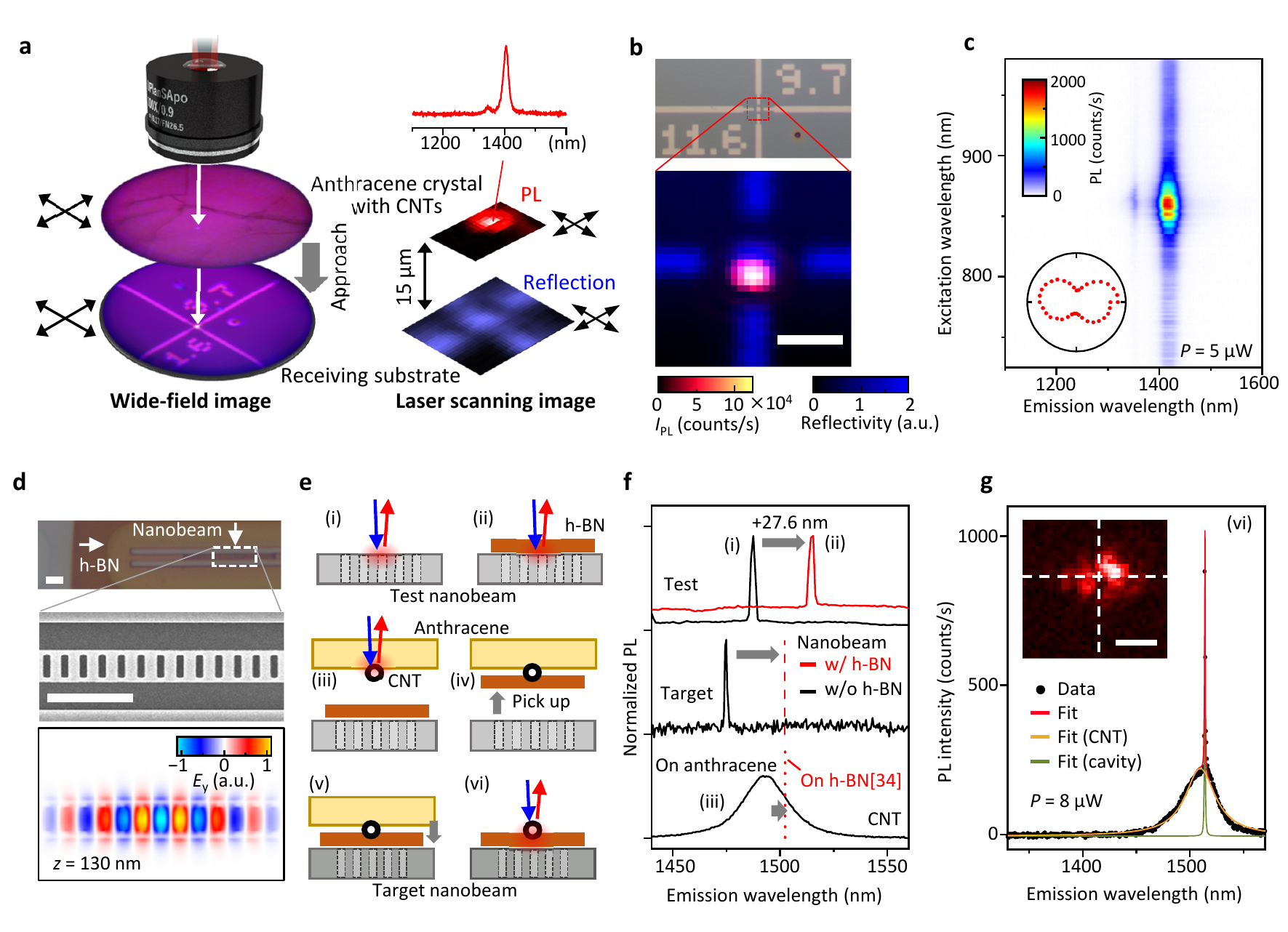}
\caption{
\label{Fig4} Demonstration of deterministic transfer of CNTs. (a) Schematic showing the alignment of a selected CNT and a receiving substrate. Wide-field images and laser scanning images are simultaneously obtained for the stamp and the substrate. (b) Optical image of the receiving substrate, which is covered by a polystyrene film to suppress exciton quenching (see Figure~\ref{Fig3}). The numbers (11,6) are lithographically patterned metal, and represent the chiral index of a CNT to be transferred (top). Confocal PL image of the transferred (11,6) nanotube (bottom), where $I_\mathrm{PL}$ at 1418~nm and the laser reflectivity are superimposed. (c) PLE map of the transferred CNT. Inset: Polarization dependence of $I_\mathrm{PL}$. Data in b and c are taken with $P=5~\mu$W. (d) Optical micrograph of a nanobeam cavity covered with a $\sim$30-nm-thick h-BN flake. Typical SEM image of an as-fabricated nanobeam cavity (middle) and simulated spatial distribution of the $y$-component of the electric field for the fundamental mode at the top surface (bottom). (e) Schematics showing the steps for measurements and deterministic transfer of a CNT/h-BN heterostructure onto a nanobeam cavity. (f) PL spectra of the fundamental mode of cavities with (red) and without (black) the h-BN nano-spacer, and PL spectrum of a (13,5) nanotube picked up on an anthracene crystal. For the test nanobeam, the resonant wavelength is redshifted by 27.6~nm due to the covering dielectric layer. Based on the shift value, a target nanobeam is chosen so that wavelengths of the shifted fundamental mode and the (13,5) tube emission coincide at $\sim$1502~nm~\cite{Fang:2020}. (g) PL spectrum of the cavity-coupled  CNT. Black dots are data and solid lines are the Lorentzian multi-peak fits. $P=8~\mu$W and laser polarization is perpendicular to the nanobeam. Inset: PL image of the (13,5) tube emission. White dashed cross indicates the center of the cavity. All the scale bars are 2~$\mu$m.
}
\end{figure*}

The 100\% release of CNTs from the stamp to the target substrate combined with the bright PL on anthracene facilitates deterministic transfer of CNTs where chirality and position are controlled simultaneously. Position-controlled transfer approaches have been limited for CNTs~\cite{Huang:2005, Waissman:2013} because individual CNTs are barely visible by conventional optical means. Electron microscopy can visualize CNTs, but it requires elaborate vacuum systems and induces exciton quenching sites~\cite{Suzuki:2007}. Since PL spectroscopy can reveal the location, angle, and chirality of single CNTs, we have incorporated the CNT transfer system into the micro-PL setup. Hyperspectral 2D images and PLE maps are obtained to locate a CNT with a desired chirality among CNTs on anthracene crystal surface. As shown in Figure~\ref{Fig4}(a), we align the selected CNT and the destination on the receiving substrate to the optical axis of the laser beam. The anthracene crystal with the CNT is pressed and then releases at the designated position.

We demonstrate such a deterministic approach by transferring the CNTs, whose chiral indices are pre-specified on a receiving substrate as shown in the upper part of Figure~\ref{Fig4}(b). A nanotube with chirality of (11,6) is found on the anthracene crystal and then transferred to the center of the cross mark. PL signals at $\lambda_\mathrm{em}= 1418$~nm and the laser reflectivity are superimposed (Figure~\ref{Fig4}(b), bottom), displaying the desired CNT at the designated location with only $\sim$500~nm misalignment. The PLE map confirms the tube chirality to be (11,6) (Figure~\ref{Fig4}(c)), thereby demonstrating that our transfer method provides a sub-micron position accuracy and chirality selectivity for individual bright CNTs. We have subsequently transferred a (9,7) nanotube at the same location and the result is summarized in Figure~S11, further supporting the reliability of our method.

The benefits of the position and chirality control over transferred CNTs are more evident in achieving deterministic coupling of a CNT to a photonic crystal microcavity, in which amplified electric fields are localized within a few $\mu$m~\cite{Miura:2014}. We demonstrate the capability of our transfer method by precisely placing a CNT whose emission wavelength matches the fundamental mode of a nanobeam cavity. As shown in Figure~\ref{Fig4}(d), a hexagonal boron nitride (h-BN) flake covers an as-fabricated nanobeam cavity and serves as a nano-spacer to simultaneously prevent exciton quenching and the decay of electric fields. 

We perform deterministic transfer of a CNT and h-BN by taking into account the spectral shifts of the CNT and the cavity modes in contact with h-BN. The fabrication steps are summarized in Figure~\ref{Fig4}(e). The fundamental mode of a test nanobeam cavity is redshifted by 27.6~nm after the transfer of an h-BN flake (Figure~\ref{Fig4}(f), top). Based on the shift amount, we seek a target nanobeam whose cavity mode under this particular h-BN flake would match the (13,5) tube emission on h-BN~\cite{Fang:2020} (Figure~\ref{Fig4}(f), middle). After characterizing a (13,5) CNT on an anthracene crystal, we further pick up the h-BN flake and place this anthracene/CNT/h-BN heterostructure onto the target nanobeam, followed by anthracene sublimation. A PL spectrum of the CNT in Figure~\ref{Fig4}(g) has a cavity-coupled narrow line, exhibiting a nearly perfect spectral and spatial matching. We note that the high compatibility of anthracene crystals with 2D materials plays a key role in this demonstration (see Figure~S9 for the case of suspended monolayer WSe$_2$).

\paragraph*{Exciton transfer at a cross junction.}

\begin{figure}
\includegraphics{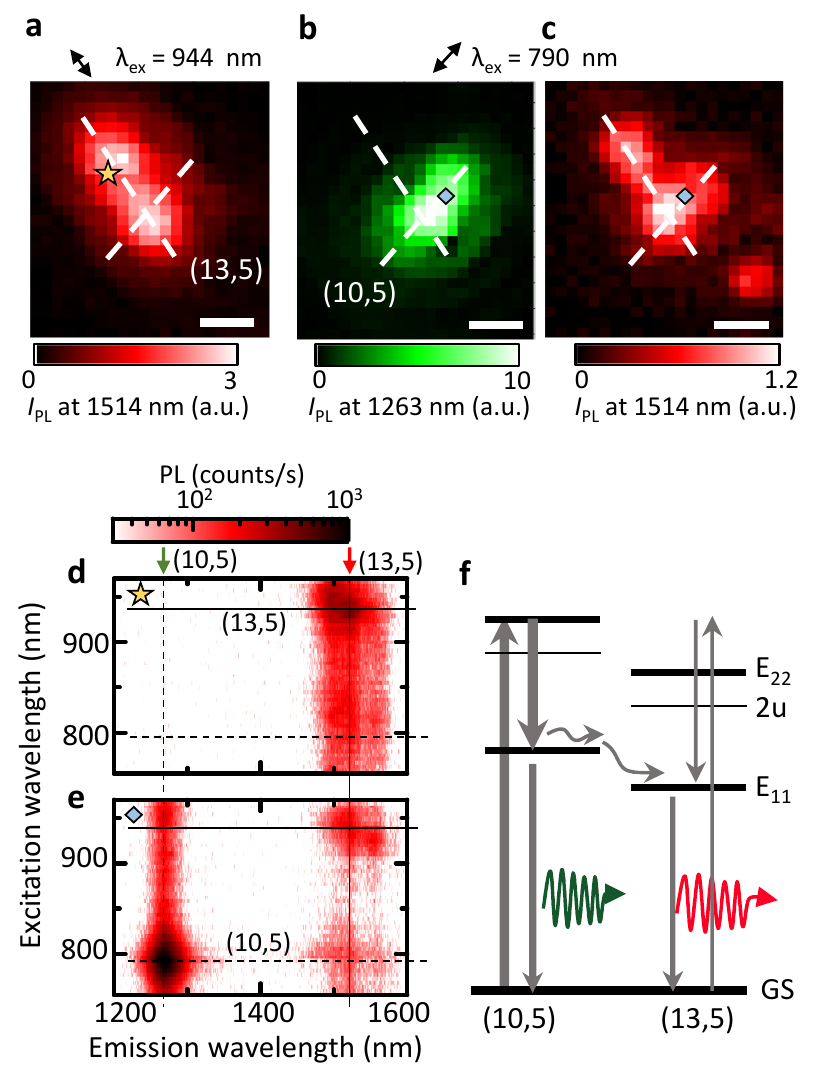}
\caption{
\label{Fig5} Crossed CNTs formed through multiple transfer processes. (a) Confocal PL image of a transferred (13,5) CNT, which has an emission peak at 1514~nm, with the excitation at its E$_{22}$ resonance (944~nm) and the laser polarization parallel to the tube axis. (b) PL image of a (10,5) tube at $\lambda_\mathrm{em}=1263$~nm, taken with excitation near its E$_{22}$ resonance (790~nm) and the polarization parallel to its axis. (c) PL image at $\lambda_\mathrm{em}=1514$~nm, which is simultaneously obtained with (b). Scale bars are 1~$\mu$m. (d,e) PLE maps measured at the yellow star and the blue diamonds in (a--c). (f) Schematic of energy diagram for the (10,5)-(13,5) CNT heterostructure, showing emission at 1263 and 1514~nm with the excitation at 790~nm. $P=100~\mu$W for all the panels. The substrate is covered with a PMMA film before the CNT transfer, and measurements are performed without removing the anthracene crystal used for transferring the (13,5) tube.
}
\end{figure}

Our transfer technique also offers opportunities to study basic elements needed to build advanced structures consisting of multiple CNTs. Cross junction formation is an example, which allows us to extract intertube exciton dynamics at controlled interfaces. A PL image and a PLE map of an isolated (10,5) tube that is transferred first are shown in Figure~S12. Next, we find a (13,5) tube among the CNTs picked up from the quartz substrate, whose angle is adjusted perpendicular to the pre-transferred (10,5) tube. The (13,5) tube and the anthracene crystal are then released over the (10,5) nanotube on the chip.

We perform PL imaging with two different excitation conditions. Figure~\ref{Fig5}(a) shows $I_\mathrm{PL}$ from the (13,5) tube when the excitation energy and polarization are resonant and parallel, respectively, showing a straight geometry of the transferred CNT. When the laser wavelength and polarization are chosen to maximize the $I_\mathrm{PL}$ of the (10,5) tube, the PL image for the (10,5) tube emission does not show any remarkable change before and after the transfer of the additional tube (Figures~S12(a) and \ref{Fig5}(b)). Interestingly, the PL image filtered around the (13,5) tube emission peak (Figure~\ref{Fig5}(c)) shows a small bump that spatially overlaps with the (10,5) tube. This implies that higher energy excitons generated in the thinner (10,5) tube diffuse towards the cross junction before transferring into the other nanotube. 

We also observe spectral signatures of intertube exciton transfer in PLE spectra. The PLE map in Figure~\ref{Fig5}(d) is measured at the yellow star in Figure~\ref{Fig5}(a), showing the E$_\mathrm{22}$ absorption peak of the (13,5) tube without other prominent absorption peaks. In contrast, when tubes are excited at the light blue diamond in Figure~\ref{Fig5}(b,c), the PLE spectrum for the (13,5) tube emission shows a sub-peak at the E$_\mathrm{22}$ absorption peak of the (10,5) tube (Figure~\ref{Fig5}(e)). This absorption peak for the (13,5) tube emission indicates that a small portion of excitons generated in the (10,5) tube is transferred and contributes to the emission around 1500~nm (Figure~\ref{Fig5}(f)). The excitonic energy mismatch between the CNTs, as well as the 0D-like ultrasmall contact, may lead to the low efficiency of exciton transfer of $\sim$8\% (see Figure~S12) compared with a previous study~\cite{Qian:2008}. 

\section{Conclusions}
We have developed a method for deterministic transfer of chirality-on-demand CNTs with high optical quality, assisted by sacrificial anthracene single crystals. Transferred CNTs fluoresce as bright as unprocessed air-suspended CNTs, where crystal growth and removal by sublimation play a key role. With precise control over position, angle, and tube chirality, individual nanotubes are coupled in a deterministic manner to other nanoscale components such as a nanobeam cavity and another nanotube. Our results demonstrate the extensive capabilities of the deterministic transfer method in nanodevice fabrication from atomically defined building blocks. Novel materials and automated fabrication systems would allow for larger scale and higher complexity structures, eventually leading to technology that make use of all the room at the bottom.

\section*{Methods}
\paragraph*{Micro-PL measurements.}
A homebuilt confocal microscopy system is used to perform PL measurements at room temperature in air. We use a wavelength-tunable Ti:sapphire laser for excitation with its power and polarization angle controlled by neutral density filters and a half-wave plate, respectively. The laser beam is focused on the samples using an objective lens with a correction collar, a numerical aperture of 0.65, and a working distance of 4.5~mm. PL is collected through the same objective lens and detected using a liquid-nitrogen-cooled \\*{InGaAs} diode array attached to a spectrometer. To assist the transfer process while monitoring PL signals, wide field 2D images of the substrate and the stamp are simultaneously obtained with a commercially available CMOS camera under coaxial illumination by an LED (see Figure~S1).

\paragraph*{Carbon nanotube synthesis.}
We use three types of CNT synthesis in this study. The arrays of long CNTs are grown on single-crystalline quartz substrates~\cite{Otsuka:2018}. Iron catalyst nanoparticles with a nominal thickness of 0.2~nm are deposited on lithographically patterned stripes. The CNT arrays are grown at 800$^\circ$C for 10~min by low-pressure alcoholic chemical vapor deposition (CVD) technique. Air-suspended CNTs are grown across 1- to 2-$\mu$m-wide trenches that are formed on SiO$_2$/Si substrates by dry etching. Iron catalyst (0.1~nm~thick) is deposited near the trenches by electron-beam evaporation, followed by atmospheric pressure CVD (1~min) at 800$^\circ$C with ethanol precursor supplied through bubbling of Ar/H$_2$ carrier gas~\cite{Ishii:2015}. For randomly oriented CNTs that are transferred via anthracene, the CNTs are grown on flat SiO$_2$/Si substrates without trench formation nor catalyst patterning, and the other growth conditions are the same as the air-suspended CNTs. 

\paragraph*{Anthracene crystal growth.}
Anthracene crystals are grown by in-air sublimation~\cite{Ye:2018}. Anthracene powder is placed on a glass slide that is heated to 80$^\circ$C, and another glass slide is placed above the anthracene source typically with a 1~mm spacing. Thin and large-area single crystals start growing on the glass surface but extend out of plane of the glass surface. To efficiently grow such thin and large-area crystals, glass slides are patterned with ink of commercially available markers, on which the nucleation of three-dimensional crystals is suppressed. Typical growth time is 10~h.

\paragraph*{Transferred CNT samples for PL comparison.}
In Figure~\ref{Fig3}(a), transferred CNTs (shown in Roman types) are prepared using the anthracene-assisted method. Polymer (PMMA and polystyrene) thin films are formed by spin-coating on Si chips with alignment markers, followed by baking at 170$^\circ$C. CNTs on anthracene are measured at suspended parts of anthracene/CNTs over square pits on SiO$_2$/Si chips (see Figure~\ref{Fig1}(j)). All the transferred samples measured in Figure~\ref{Fig3} and Figure~S6 are grown from unpatterned catalysts on flat SiO$_2$/Si substrates. For air-suspended CNTs, we scan the excitation laser beam along trenches with $\lambda_\mathrm{ex}=780$~nm, while 2D raster scan with excitation at $\lambda_\mathrm{ex}=790$~nm is performed for the rest of samples to adjust for the screening-induced energy shifts~\cite{Ohno:2006prb}. We acquire the PL signal of CNTs on quartz without a grating to increase the signal-to-noise ratio and calibrate the intensity by taking into account the grating efficiency to compare $I_\mathrm{PL}$ with other samples.

\paragraph*{Deterministic coupling with nanobeam cavities}
Air-mode nanobeam cavities are fabricated using silicon-on-insulator wafers with a top Si thickness of 260~nm and a buried oxide thickness of 1~$\mu$m. The nanobeams are patterned by electron beam lithography, followed by inductively-coupled plasma etching of the top Si layer. The buried oxide layer under the nanobeam structures is then etched by hydrofluoric acid. To align the wavelengths of a transferred CNT and a nanobeam cavity with h-BN in between, cavity modes of the nanobeams are measured using Si PL with $\lambda_\mathrm{ex}=780$~nm and $P=200~\mu$W.

\begin{acknowledgments}
Parts of this study are supported by JSPS (KAKENHI JP20H02558, JP19J00894, JP20K15137), MIC (SCOPE 191503001), and MEXT (Nanotechnology Platform JPMXP09F19UT0075). Growth of hexagonal boron nitride crystals is supported by the Element Strategy Initiative conducted by the MEXT (JPMXP0112101001), JSPS (KAKENHI JP20H00354), and JST (CREST JPMJCR15F3). K.O. and D.Y. are supported by JSPS (Research Fellowship for Young Scientists). N.F. is supported by RIKEN Special Postdoctoral Researcher Program. We thank T. Inoue, S. Maruyama, and K. Nagashio for help in the preparation of CNTs and 2D materials, as well as the Advanced Manufacturing Support Team at RIKEN for technical assistance.
\end{acknowledgments}

\section*{Author Contributions}
K.O. developed the nanotube transfer system, and performed all the nanotube-related experiments and data analysis. N.F. performed transfer and spectroscopy of 2D materials. D.Y. performed simulation, fabrication, and measurements of nanobeam cavities. T.T. and K.W. provided boron nitride crystals. Y.K.K. supervised the project. K.O. and Y.K.K. co-wrote the manuscript. All the authors commented on the manuscript.


\begin{thebibliography}{39}%
\makeatletter
\providecommand \@ifxundefined [1]{%
 \@ifx{#1\undefined}
}%
\providecommand \@ifnum [1]{%
 \ifnum #1\expandafter \@firstoftwo
 \else \expandafter \@secondoftwo
 \fi
}%
\providecommand \@ifx [1]{%
 \ifx #1\expandafter \@firstoftwo
 \else \expandafter \@secondoftwo
 \fi
}%
\providecommand \natexlab [1]{#1}%
\providecommand \enquote  [1]{``#1''}%
\providecommand \bibnamefont  [1]{#1}%
\providecommand \bibfnamefont [1]{#1}%
\providecommand \citenamefont [1]{#1}%
\providecommand \href@noop [0]{\@secondoftwo}%
\providecommand \href [0]{\begingroup \@sanitize@url \@href}%
\providecommand \@href[1]{\@@startlink{#1}\@@href}%
\providecommand \@@href[1]{\endgroup#1\@@endlink}%
\providecommand \@sanitize@url [0]{\catcode `\\12\catcode `\$12\catcode
  `\&12\catcode `\#12\catcode `\^12\catcode `\_12\catcode `\%12\relax}%
\providecommand \@@startlink[1]{}%
\providecommand \@@endlink[0]{}%
\providecommand \url  [0]{\begingroup\@sanitize@url \@url }%
\providecommand \@url [1]{\endgroup\@href {#1}{\urlprefix }}%
\providecommand \urlprefix  [0]{URL }%
\providecommand \Eprint [0]{\href }%
\providecommand \doibase [0]{http://dx.doi.org/}%
\providecommand \selectlanguage [0]{\@gobble}%
\providecommand \bibinfo  [0]{\@secondoftwo}%
\providecommand \bibfield  [0]{\@secondoftwo}%
\providecommand \translation [1]{[#1]}%
\providecommand \BibitemOpen [0]{}%
\providecommand \bibitemStop [0]{}%
\providecommand \bibitemNoStop [0]{.\EOS\space}%
\providecommand \EOS [0]{\spacefactor3000\relax}%
\providecommand \BibitemShut  [1]{\csname bibitem#1\endcsname}%
\let\auto@bib@innerbib\@empty
\bibitem [{\citenamefont {Feynman}(1960)}]{Feynman:1960}%
  \BibitemOpen
  \bibfield  {author} {\bibinfo {author} {\bibfnamefont {R.~P.}\ \bibnamefont
  {Feynman}},\ }\bibfield  {title} {\bibinfo {title} {There’s plenty of room
  at the bottom},\ }\href@noop {} {\bibfield  {journal} {\bibinfo  {journal}
  {Engineering and Science}\ }\textbf {\bibinfo {volume} {23}},\ \bibinfo
  {pages} {22} (\bibinfo {year} {1960})}\BibitemShut {NoStop}%
\bibitem [{\citenamefont {Novoselov}\ \emph {et~al.}(2004)\citenamefont
  {Novoselov}, \citenamefont {Geim}, \citenamefont {Morozov}, \citenamefont
  {Jiang}, \citenamefont {Zhang}, \citenamefont {Dubonos}, \citenamefont
  {Grigorieva},\ and\ \citenamefont {Firsov}}]{Novoselov:2004}%
  \BibitemOpen
  \bibfield  {author} {\bibinfo {author} {\bibfnamefont {K.~S.}\ \bibnamefont
  {Novoselov}}, \bibinfo {author} {\bibfnamefont {A.~K.}\ \bibnamefont {Geim}},
  \bibinfo {author} {\bibfnamefont {S.~V.}\ \bibnamefont {Morozov}}, \bibinfo
  {author} {\bibfnamefont {D.}~\bibnamefont {Jiang}}, \bibinfo {author}
  {\bibfnamefont {Y.}~\bibnamefont {Zhang}}, \bibinfo {author} {\bibfnamefont
  {S.~V.}\ \bibnamefont {Dubonos}}, \bibinfo {author} {\bibfnamefont {I.~V.}\
  \bibnamefont {Grigorieva}}, \ and\ \bibinfo {author} {\bibfnamefont {A.~A.}\
  \bibnamefont {Firsov}},\ }\bibfield  {title} {\bibinfo {title} {Electric
  field effect in atomically thin carbon films},\ }\href {\doibase
  10.1126/science.1102896} {\bibfield  {journal} {\bibinfo  {journal}
  {Science}\ }\textbf {\bibinfo {volume} {306}},\ \bibinfo {pages} {666}
  (\bibinfo {year} {2004})}\BibitemShut {NoStop}%
\bibitem [{\citenamefont {Novoselov}\ \emph {et~al.}(2005)\citenamefont
  {Novoselov}, \citenamefont {Jiang}, \citenamefont {Schedin}, \citenamefont
  {Booth}, \citenamefont {Khotkevich}, \citenamefont {Morozov},\ and\
  \citenamefont {Geim}}]{Novoselov:2005}%
  \BibitemOpen
  \bibfield  {author} {\bibinfo {author} {\bibfnamefont {K.~S.}\ \bibnamefont
  {Novoselov}}, \bibinfo {author} {\bibfnamefont {D.}~\bibnamefont {Jiang}},
  \bibinfo {author} {\bibfnamefont {F.}~\bibnamefont {Schedin}}, \bibinfo
  {author} {\bibfnamefont {T.~J.}\ \bibnamefont {Booth}}, \bibinfo {author}
  {\bibfnamefont {V.~V.}\ \bibnamefont {Khotkevich}}, \bibinfo {author}
  {\bibfnamefont {S.~V.}\ \bibnamefont {Morozov}}, \ and\ \bibinfo {author}
  {\bibfnamefont {A.~K.}\ \bibnamefont {Geim}},\ }\bibfield  {title} {\bibinfo
  {title} {Two-dimensional atomic crystals},\ }\href {\doibase
  10.1073/pnas.0502848102} {\bibfield  {journal} {\bibinfo  {journal} {Proc.
  Natl. Acad. Sci. USA}\ }\textbf {\bibinfo {volume} {102}},\ \bibinfo {pages}
  {10451} (\bibinfo {year} {2005})}\BibitemShut {NoStop}%
\bibitem [{\citenamefont {Dean}\ \emph {et~al.}(2010)\citenamefont {Dean},
  \citenamefont {Young}, \citenamefont {Meric}, \citenamefont {Lee},
  \citenamefont {Wang}, \citenamefont {Sorgenfrei}, \citenamefont {Watanabe},
  \citenamefont {Taniguchi}, \citenamefont {Kim}, \citenamefont {Shepard},\
  and\ \citenamefont {Hone}}]{Dean:2010}%
  \BibitemOpen
  \bibfield  {author} {\bibinfo {author} {\bibfnamefont {C.~R.}\ \bibnamefont
  {Dean}}, \bibinfo {author} {\bibfnamefont {A.~F.}\ \bibnamefont {Young}},
  \bibinfo {author} {\bibfnamefont {I.}~\bibnamefont {Meric}}, \bibinfo
  {author} {\bibfnamefont {C.}~\bibnamefont {Lee}}, \bibinfo {author}
  {\bibfnamefont {L.}~\bibnamefont {Wang}}, \bibinfo {author} {\bibfnamefont
  {S.}~\bibnamefont {Sorgenfrei}}, \bibinfo {author} {\bibfnamefont
  {K.}~\bibnamefont {Watanabe}}, \bibinfo {author} {\bibfnamefont
  {T.}~\bibnamefont {Taniguchi}}, \bibinfo {author} {\bibfnamefont
  {P.}~\bibnamefont {Kim}}, \bibinfo {author} {\bibfnamefont {K.~L.}\
  \bibnamefont {Shepard}}, \ and\ \bibinfo {author} {\bibfnamefont
  {J.}~\bibnamefont {Hone}},\ }\bibfield  {title} {\bibinfo {title} {Boron
  nitride substrates for high-quality graphene electronics},\ }\href {\doibase
  10.1038/nnano.2010.172} {\bibfield  {journal} {\bibinfo  {journal} {Nat.
  Nanotechnol.}\ }\textbf {\bibinfo {volume} {5}},\ \bibinfo {pages} {722}
  (\bibinfo {year} {2010})}\BibitemShut {NoStop}%
\bibitem [{\citenamefont {Cao}\ \emph {et~al.}(2018)\citenamefont {Cao},
  \citenamefont {Fatemi}, \citenamefont {Fang}, \citenamefont {Watanabe},
  \citenamefont {Taniguchi}, \citenamefont {Kaxiras},\ and\ \citenamefont
  {Jarillo-Herrero}}]{Cao:2018}%
  \BibitemOpen
  \bibfield  {author} {\bibinfo {author} {\bibfnamefont {Y.}~\bibnamefont
  {Cao}}, \bibinfo {author} {\bibfnamefont {V.}~\bibnamefont {Fatemi}},
  \bibinfo {author} {\bibfnamefont {S.}~\bibnamefont {Fang}}, \bibinfo {author}
  {\bibfnamefont {K.}~\bibnamefont {Watanabe}}, \bibinfo {author}
  {\bibfnamefont {T.}~\bibnamefont {Taniguchi}}, \bibinfo {author}
  {\bibfnamefont {E.}~\bibnamefont {Kaxiras}}, \ and\ \bibinfo {author}
  {\bibfnamefont {P.}~\bibnamefont {Jarillo-Herrero}},\ }\bibfield  {title}
  {\bibinfo {title} {Unconventional superconductivity in magic-angle graphene
  superlattices},\ }\href {\doibase 10.1038/nature26160} {\bibfield  {journal}
  {\bibinfo  {journal} {Nature}\ }\textbf {\bibinfo {volume} {556}},\ \bibinfo
  {pages} {43} (\bibinfo {year} {2018})}\BibitemShut {NoStop}%
\bibitem [{\citenamefont {Shimazaki}\ \emph {et~al.}(2020)\citenamefont
  {Shimazaki}, \citenamefont {Schwartz}, \citenamefont {Watanabe},
  \citenamefont {Taniguchi}, \citenamefont {Kroner},\ and\ \citenamefont
  {Imamo{\u{g}}lu}}]{Shimazaki:2020}%
  \BibitemOpen
  \bibfield  {author} {\bibinfo {author} {\bibfnamefont {Y.}~\bibnamefont
  {Shimazaki}}, \bibinfo {author} {\bibfnamefont {I.}~\bibnamefont {Schwartz}},
  \bibinfo {author} {\bibfnamefont {K.}~\bibnamefont {Watanabe}}, \bibinfo
  {author} {\bibfnamefont {T.}~\bibnamefont {Taniguchi}}, \bibinfo {author}
  {\bibfnamefont {M.}~\bibnamefont {Kroner}}, \ and\ \bibinfo {author}
  {\bibfnamefont {A.}~\bibnamefont {Imamo{\u{g}}lu}},\ }\bibfield  {title}
  {\bibinfo {title} {Strongly correlated electrons and hybrid excitons in a
  moir{\'{e}} heterostructure},\ }\href {\doibase 10.1038/s41586-020-2191-2}
  {\bibfield  {journal} {\bibinfo  {journal} {Nature}\ }\textbf {\bibinfo
  {volume} {580}},\ \bibinfo {pages} {472} (\bibinfo {year}
  {2020})}\BibitemShut {NoStop}%
\bibitem [{\citenamefont {Wang}\ \emph {et~al.}(2019)\citenamefont {Wang},
  \citenamefont {Rhodes}, \citenamefont {Watanabe}, \citenamefont {Taniguchi},
  \citenamefont {Hone}, \citenamefont {Shan},\ and\ \citenamefont
  {Mak}}]{Wang:2019}%
  \BibitemOpen
  \bibfield  {author} {\bibinfo {author} {\bibfnamefont {Z.}~\bibnamefont
  {Wang}}, \bibinfo {author} {\bibfnamefont {D.~A.}\ \bibnamefont {Rhodes}},
  \bibinfo {author} {\bibfnamefont {K.}~\bibnamefont {Watanabe}}, \bibinfo
  {author} {\bibfnamefont {T.}~\bibnamefont {Taniguchi}}, \bibinfo {author}
  {\bibfnamefont {J.~C.}\ \bibnamefont {Hone}}, \bibinfo {author}
  {\bibfnamefont {J.}~\bibnamefont {Shan}}, \ and\ \bibinfo {author}
  {\bibfnamefont {K.~F.}\ \bibnamefont {Mak}},\ }\bibfield  {title} {\bibinfo
  {title} {Evidence of high-temperature exciton condensation in two-dimensional
  atomic double layers},\ }\href {\doibase 10.1038/s41586-019-1591-7}
  {\bibfield  {journal} {\bibinfo  {journal} {Nature}\ }\textbf {\bibinfo
  {volume} {574}},\ \bibinfo {pages} {76} (\bibinfo {year} {2019})}\BibitemShut
  {NoStop}%
\bibitem [{\citenamefont {Park}\ \emph {et~al.}(2000)\citenamefont {Park},
  \citenamefont {Park}, \citenamefont {Lim}, \citenamefont {Anderson},
  \citenamefont {Alivisatos},\ and\ \citenamefont {McEuen}}]{Park:2000}%
  \BibitemOpen
  \bibfield  {author} {\bibinfo {author} {\bibfnamefont {H.}~\bibnamefont
  {Park}}, \bibinfo {author} {\bibfnamefont {J.}~\bibnamefont {Park}}, \bibinfo
  {author} {\bibfnamefont {A.~K.~L.}\ \bibnamefont {Lim}}, \bibinfo {author}
  {\bibfnamefont {E.~H.}\ \bibnamefont {Anderson}}, \bibinfo {author}
  {\bibfnamefont {A.~P.}\ \bibnamefont {Alivisatos}}, \ and\ \bibinfo {author}
  {\bibfnamefont {P.~L.}\ \bibnamefont {McEuen}},\ }\bibfield  {title}
  {\bibinfo {title} {Nanomechanical oscillations in a single-{C$_{60}$}
  transistor},\ }\href {\doibase 10.1038/35024031} {\bibfield  {journal}
  {\bibinfo  {journal} {Nature}\ }\textbf {\bibinfo {volume} {407}},\ \bibinfo
  {pages} {57} (\bibinfo {year} {2000})}\BibitemShut {NoStop}%
\bibitem [{\citenamefont {Perrin}\ \emph {et~al.}(2015)\citenamefont {Perrin},
  \citenamefont {Burzur{\'{\i}}},\ and\ \citenamefont {van~der
  Zant}}]{Perrin:2015}%
  \BibitemOpen
  \bibfield  {author} {\bibinfo {author} {\bibfnamefont {M.~L.}\ \bibnamefont
  {Perrin}}, \bibinfo {author} {\bibfnamefont {E.}~\bibnamefont
  {Burzur{\'{\i}}}}, \ and\ \bibinfo {author} {\bibfnamefont {H.~S.~J.}\
  \bibnamefont {van~der Zant}},\ }\bibfield  {title} {\bibinfo {title}
  {Single-molecule transistors},\ }\href {\doibase 10.1039/c4cs00231h}
  {\bibfield  {journal} {\bibinfo  {journal} {Chem. Soc. Rev}\ }\textbf
  {\bibinfo {volume} {44}},\ \bibinfo {pages} {902} (\bibinfo {year}
  {2015})}\BibitemShut {NoStop}%
\bibitem [{\citenamefont {Moresco}\ \emph {et~al.}(2001)\citenamefont
  {Moresco}, \citenamefont {Meyer}, \citenamefont {Rieder}, \citenamefont
  {Tang}, \citenamefont {Gourdon},\ and\ \citenamefont
  {Joachim}}]{Moresco:2001}%
  \BibitemOpen
  \bibfield  {author} {\bibinfo {author} {\bibfnamefont {F.}~\bibnamefont
  {Moresco}}, \bibinfo {author} {\bibfnamefont {G.}~\bibnamefont {Meyer}},
  \bibinfo {author} {\bibfnamefont {K.-H.}\ \bibnamefont {Rieder}}, \bibinfo
  {author} {\bibfnamefont {H.}~\bibnamefont {Tang}}, \bibinfo {author}
  {\bibfnamefont {A.}~\bibnamefont {Gourdon}}, \ and\ \bibinfo {author}
  {\bibfnamefont {C.}~\bibnamefont {Joachim}},\ }\bibfield  {title} {\bibinfo
  {title} {Conformational changes of single molecules induced by scanning
  tunneling microscopy manipulation: A route to molecular switching},\ }\href
  {\doibase 10.1103/physrevlett.86.672} {\bibfield  {journal} {\bibinfo
  {journal} {Phys. Rev. Lett.}\ }\textbf {\bibinfo {volume} {86}},\ \bibinfo
  {pages} {672} (\bibinfo {year} {2001})}\BibitemShut {NoStop}%
\bibitem [{\citenamefont {Klein}\ \emph {et~al.}(1997)\citenamefont {Klein},
  \citenamefont {Roth}, \citenamefont {Lim}, \citenamefont {Alivisatos},\ and\
  \citenamefont {McEuen}}]{Klein:1997}%
  \BibitemOpen
  \bibfield  {author} {\bibinfo {author} {\bibfnamefont {D.~L.}\ \bibnamefont
  {Klein}}, \bibinfo {author} {\bibfnamefont {R.}~\bibnamefont {Roth}},
  \bibinfo {author} {\bibfnamefont {A.~K.~L.}\ \bibnamefont {Lim}}, \bibinfo
  {author} {\bibfnamefont {A.~P.}\ \bibnamefont {Alivisatos}}, \ and\ \bibinfo
  {author} {\bibfnamefont {P.~L.}\ \bibnamefont {McEuen}},\ }\bibfield  {title}
  {\bibinfo {title} {A single-electron transistor made from a cadmium selenide
  nanocrystal},\ }\href {\doibase 10.1038/39535} {\bibfield  {journal}
  {\bibinfo  {journal} {Nature}\ }\textbf {\bibinfo {volume} {389}},\ \bibinfo
  {pages} {699} (\bibinfo {year} {1997})}\BibitemShut {NoStop}%
\bibitem [{\citenamefont {Santori}\ \emph {et~al.}(2002)\citenamefont
  {Santori}, \citenamefont {Fattal}, \citenamefont {Vu{\v{c}}kovi{\'{c}}},
  \citenamefont {Solomon},\ and\ \citenamefont {Yamamoto}}]{Santori:2002}%
  \BibitemOpen
  \bibfield  {author} {\bibinfo {author} {\bibfnamefont {C.}~\bibnamefont
  {Santori}}, \bibinfo {author} {\bibfnamefont {D.}~\bibnamefont {Fattal}},
  \bibinfo {author} {\bibfnamefont {J.}~\bibnamefont {Vu{\v{c}}kovi{\'{c}}}},
  \bibinfo {author} {\bibfnamefont {G.~S.}\ \bibnamefont {Solomon}}, \ and\
  \bibinfo {author} {\bibfnamefont {Y.}~\bibnamefont {Yamamoto}},\ }\bibfield
  {title} {\bibinfo {title} {Indistinguishable photons from a single-photon
  device},\ }\href {\doibase 10.1038/nature01086} {\bibfield  {journal}
  {\bibinfo  {journal} {Nature}\ }\textbf {\bibinfo {volume} {419}},\ \bibinfo
  {pages} {594} (\bibinfo {year} {2002})}\BibitemShut {NoStop}%
\bibitem [{\citenamefont {Postma}\ \emph {et~al.}(2001)\citenamefont {Postma},
  \citenamefont {Teepen}, \citenamefont {Yao}, \citenamefont {Grifoni},\ and\
  \citenamefont {Dekker}}]{Postma:2001}%
  \BibitemOpen
  \bibfield  {author} {\bibinfo {author} {\bibfnamefont {H.~W.~C.}\
  \bibnamefont {Postma}}, \bibinfo {author} {\bibfnamefont {T.}~\bibnamefont
  {Teepen}}, \bibinfo {author} {\bibfnamefont {Z.}~\bibnamefont {Yao}},
  \bibinfo {author} {\bibfnamefont {M.}~\bibnamefont {Grifoni}}, \ and\
  \bibinfo {author} {\bibfnamefont {C.}~\bibnamefont {Dekker}},\ }\bibfield
  {title} {\bibinfo {title} {Carbon nanotube single-electron transistors at
  room temperature},\ }\href {\doibase 10.1126/science.1061797} {\bibfield
  {journal} {\bibinfo  {journal} {Science}\ }\textbf {\bibinfo {volume}
  {293}},\ \bibinfo {pages} {76} (\bibinfo {year} {2001})}\BibitemShut
  {NoStop}%
\bibitem [{\citenamefont {Mueller}\ \emph {et~al.}(2010)\citenamefont
  {Mueller}, \citenamefont {Kinoshita}, \citenamefont {Steiner}, \citenamefont
  {Perebeinos}, \citenamefont {Bol}, \citenamefont {Farmer},\ and\
  \citenamefont {Avouris}}]{Mueller:2010}%
  \BibitemOpen
  \bibfield  {author} {\bibinfo {author} {\bibfnamefont {T.}~\bibnamefont
  {Mueller}}, \bibinfo {author} {\bibfnamefont {M.}~\bibnamefont {Kinoshita}},
  \bibinfo {author} {\bibfnamefont {M.}~\bibnamefont {Steiner}}, \bibinfo
  {author} {\bibfnamefont {V.}~\bibnamefont {Perebeinos}}, \bibinfo {author}
  {\bibfnamefont {A.~A.}\ \bibnamefont {Bol}}, \bibinfo {author} {\bibfnamefont
  {D.~B.}\ \bibnamefont {Farmer}}, \ and\ \bibinfo {author} {\bibfnamefont
  {P.}~\bibnamefont {Avouris}},\ }\bibfield  {title} {\bibinfo {title}
  {Efficient narrow-band light emission from a single carbon nanotube p-n
  diode},\ }\href {\doibase 10.1038/nnano.2009.319} {\bibfield  {journal}
  {\bibinfo  {journal} {Nat. Nanotech.}\ }\textbf {\bibinfo {volume} {5}},\
  \bibinfo {pages} {27} (\bibinfo {year} {2010})}\BibitemShut {NoStop}%
\bibitem [{\citenamefont {Huang}\ \emph {et~al.}(2005)\citenamefont {Huang},
  \citenamefont {Caldwell}, \citenamefont {Huang}, \citenamefont {Jun},
  \citenamefont {Huang}, \citenamefont {Sfeir}, \citenamefont
  {O{\textquotesingle}Brien},\ and\ \citenamefont {Hone}}]{Huang:2005}%
  \BibitemOpen
  \bibfield  {author} {\bibinfo {author} {\bibfnamefont {X.~M.~H.}\
  \bibnamefont {Huang}}, \bibinfo {author} {\bibfnamefont {R.}~\bibnamefont
  {Caldwell}}, \bibinfo {author} {\bibfnamefont {L.}~\bibnamefont {Huang}},
  \bibinfo {author} {\bibfnamefont {S.~C.}\ \bibnamefont {Jun}}, \bibinfo
  {author} {\bibfnamefont {M.}~\bibnamefont {Huang}}, \bibinfo {author}
  {\bibfnamefont {M.~Y.}\ \bibnamefont {Sfeir}}, \bibinfo {author}
  {\bibfnamefont {S.~P.}\ \bibnamefont {O{\textquotesingle}Brien}}, \ and\
  \bibinfo {author} {\bibfnamefont {J.}~\bibnamefont {Hone}},\ }\bibfield
  {title} {\bibinfo {title} {Controlled placement of individual carbon
  nanotubes},\ }\href {\doibase 10.1021/nl050886a} {\bibfield  {journal}
  {\bibinfo  {journal} {Nano Lett.}\ }\textbf {\bibinfo {volume} {5}},\
  \bibinfo {pages} {1515} (\bibinfo {year} {2005})}\BibitemShut {NoStop}%
\bibitem [{\citenamefont {Abrams}\ \emph {et~al.}(2007)\citenamefont {Abrams},
  \citenamefont {Ioffe}, \citenamefont {Tsukernik}, \citenamefont
  {Cheshnovsky},\ and\ \citenamefont {Hanein}}]{Abrams:2007}%
  \BibitemOpen
  \bibfield  {author} {\bibinfo {author} {\bibfnamefont {Z.~R.}\ \bibnamefont
  {Abrams}}, \bibinfo {author} {\bibfnamefont {Z.}~\bibnamefont {Ioffe}},
  \bibinfo {author} {\bibfnamefont {A.}~\bibnamefont {Tsukernik}}, \bibinfo
  {author} {\bibfnamefont {O.}~\bibnamefont {Cheshnovsky}}, \ and\ \bibinfo
  {author} {\bibfnamefont {Y.}~\bibnamefont {Hanein}},\ }\bibfield  {title}
  {\bibinfo {title} {A complete scheme for creating predefined networks of
  individual carbon nanotubes},\ }\href {\doibase 10.1021/nl071058f} {\bibfield
   {journal} {\bibinfo  {journal} {Nano Lett.}\ }\textbf {\bibinfo {volume}
  {7}},\ \bibinfo {pages} {2666} (\bibinfo {year} {2007})}\BibitemShut
  {NoStop}%
\bibitem [{\citenamefont {Wu}\ \emph {et~al.}(2010)\citenamefont {Wu},
  \citenamefont {Liu},\ and\ \citenamefont {Zhong}}]{Wu:2010}%
  \BibitemOpen
  \bibfield  {author} {\bibinfo {author} {\bibfnamefont {C.~C.}\ \bibnamefont
  {Wu}}, \bibinfo {author} {\bibfnamefont {C.~H.}\ \bibnamefont {Liu}}, \ and\
  \bibinfo {author} {\bibfnamefont {Z.}~\bibnamefont {Zhong}},\ }\bibfield
  {title} {\bibinfo {title} {One-step direct transfer of pristine single-walled
  carbon nanotubes for functional nanoelectronics},\ }\href {\doibase
  10.1021/nl904260k} {\bibfield  {journal} {\bibinfo  {journal} {Nano Lett.}\
  }\textbf {\bibinfo {volume} {10}},\ \bibinfo {pages} {1032} (\bibinfo {year}
  {2010})}\BibitemShut {NoStop}%
\bibitem [{\citenamefont {Waissman}\ \emph {et~al.}(2013)\citenamefont
  {Waissman}, \citenamefont {Honig}, \citenamefont {Pecker}, \citenamefont
  {Benyamini}, \citenamefont {Hamo},\ and\ \citenamefont
  {Ilani}}]{Waissman:2013}%
  \BibitemOpen
  \bibfield  {author} {\bibinfo {author} {\bibfnamefont {J.}~\bibnamefont
  {Waissman}}, \bibinfo {author} {\bibfnamefont {M.}~\bibnamefont {Honig}},
  \bibinfo {author} {\bibfnamefont {S.}~\bibnamefont {Pecker}}, \bibinfo
  {author} {\bibfnamefont {A.}~\bibnamefont {Benyamini}}, \bibinfo {author}
  {\bibfnamefont {A.}~\bibnamefont {Hamo}}, \ and\ \bibinfo {author}
  {\bibfnamefont {S.}~\bibnamefont {Ilani}},\ }\bibfield  {title} {\bibinfo
  {title} {Realization of pristine and locally tunable one-dimensional electron
  systems in carbon nanotubes},\ }\href {\doibase 10.1038/nnano.2013.143}
  {\bibfield  {journal} {\bibinfo  {journal} {Nat. Nanotechnol.}\ }\textbf
  {\bibinfo {volume} {8}},\ \bibinfo {pages} {569} (\bibinfo {year}
  {2013})}\BibitemShut {NoStop}%
\bibitem [{\citenamefont {Kocabas}\ \emph {et~al.}(2005)\citenamefont
  {Kocabas}, \citenamefont {Hur}, \citenamefont {Gaur}, \citenamefont {Meitl},
  \citenamefont {Shim},\ and\ \citenamefont {Rogers}}]{Kocabas:2005}%
  \BibitemOpen
  \bibfield  {author} {\bibinfo {author} {\bibfnamefont {C.}~\bibnamefont
  {Kocabas}}, \bibinfo {author} {\bibfnamefont {S.-H.}\ \bibnamefont {Hur}},
  \bibinfo {author} {\bibfnamefont {A.}~\bibnamefont {Gaur}}, \bibinfo {author}
  {\bibfnamefont {M.}~\bibnamefont {Meitl}}, \bibinfo {author} {\bibfnamefont
  {M.}~\bibnamefont {Shim}}, \ and\ \bibinfo {author} {\bibfnamefont
  {J.}~\bibnamefont {Rogers}},\ }\bibfield  {title} {\bibinfo {title} {Guided
  growth of large-scale, horizontally aligned arrays of single-walled carbon
  nanotubes and their use in thin-film transistors},\ }\href {\doibase
  10.1002/smll.200500120} {\bibfield  {journal} {\bibinfo  {journal} {Small}\
  }\textbf {\bibinfo {volume} {1}},\ \bibinfo {pages} {1110} (\bibinfo {year}
  {2005})}\BibitemShut {NoStop}%
\bibitem [{\citenamefont {Yang}\ \emph {et~al.}(2014)\citenamefont {Yang},
  \citenamefont {Wang}, \citenamefont {Zhang}, \citenamefont {Yang},
  \citenamefont {Luo}, \citenamefont {Xu}, \citenamefont {Wei}, \citenamefont
  {Wang}, \citenamefont {Xu}, \citenamefont {Peng}, \citenamefont {Li},
  \citenamefont {Li}, \citenamefont {Li}, \citenamefont {Li}, \citenamefont
  {Bai}, \citenamefont {Ding},\ and\ \citenamefont {Li}}]{Yang:2014}%
  \BibitemOpen
  \bibfield  {author} {\bibinfo {author} {\bibfnamefont {F.}~\bibnamefont
  {Yang}}, \bibinfo {author} {\bibfnamefont {X.}~\bibnamefont {Wang}}, \bibinfo
  {author} {\bibfnamefont {D.}~\bibnamefont {Zhang}}, \bibinfo {author}
  {\bibfnamefont {J.}~\bibnamefont {Yang}}, \bibinfo {author} {\bibfnamefont
  {D.}~\bibnamefont {Luo}}, \bibinfo {author} {\bibfnamefont {Z.}~\bibnamefont
  {Xu}}, \bibinfo {author} {\bibfnamefont {J.}~\bibnamefont {Wei}}, \bibinfo
  {author} {\bibfnamefont {J.-Q.}\ \bibnamefont {Wang}}, \bibinfo {author}
  {\bibfnamefont {Z.}~\bibnamefont {Xu}}, \bibinfo {author} {\bibfnamefont
  {F.}~\bibnamefont {Peng}}, \bibinfo {author} {\bibfnamefont {X.}~\bibnamefont
  {Li}}, \bibinfo {author} {\bibfnamefont {R.}~\bibnamefont {Li}}, \bibinfo
  {author} {\bibfnamefont {Y.}~\bibnamefont {Li}}, \bibinfo {author}
  {\bibfnamefont {M.}~\bibnamefont {Li}}, \bibinfo {author} {\bibfnamefont
  {X.}~\bibnamefont {Bai}}, \bibinfo {author} {\bibfnamefont {F.}~\bibnamefont
  {Ding}}, \ and\ \bibinfo {author} {\bibfnamefont {Y.}~\bibnamefont {Li}},\
  }\bibfield  {title} {\bibinfo {title} {Chirality-specific growth of
  single-walled carbon nanotubes on solid alloy catalysts},\ }\href {\doibase
  10.1038/nature13434} {\bibfield  {journal} {\bibinfo  {journal} {Nature}\
  }\textbf {\bibinfo {volume} {510}},\ \bibinfo {pages} {522} (\bibinfo {year}
  {2014})}\BibitemShut {NoStop}%
\bibitem [{\citenamefont {Zhang}\ \emph {et~al.}(2017)\citenamefont {Zhang},
  \citenamefont {Kang}, \citenamefont {Wang}, \citenamefont {Tong},
  \citenamefont {Yang}, \citenamefont {Wang}, \citenamefont {Qi}, \citenamefont
  {Deng}, \citenamefont {Li}, \citenamefont {Bai}, \citenamefont {Ding},\ and\
  \citenamefont {Zhang}}]{Zhang:2017nature}%
  \BibitemOpen
  \bibfield  {author} {\bibinfo {author} {\bibfnamefont {S.}~\bibnamefont
  {Zhang}}, \bibinfo {author} {\bibfnamefont {L.}~\bibnamefont {Kang}},
  \bibinfo {author} {\bibfnamefont {X.}~\bibnamefont {Wang}}, \bibinfo {author}
  {\bibfnamefont {L.}~\bibnamefont {Tong}}, \bibinfo {author} {\bibfnamefont
  {L.}~\bibnamefont {Yang}}, \bibinfo {author} {\bibfnamefont {Z.}~\bibnamefont
  {Wang}}, \bibinfo {author} {\bibfnamefont {K.}~\bibnamefont {Qi}}, \bibinfo
  {author} {\bibfnamefont {S.}~\bibnamefont {Deng}}, \bibinfo {author}
  {\bibfnamefont {Q.}~\bibnamefont {Li}}, \bibinfo {author} {\bibfnamefont
  {X.}~\bibnamefont {Bai}}, \bibinfo {author} {\bibfnamefont {F.}~\bibnamefont
  {Ding}}, \ and\ \bibinfo {author} {\bibfnamefont {J.}~\bibnamefont {Zhang}},\
  }\bibfield  {title} {\bibinfo {title} {Arrays of horizontal carbon nanotubes
  of controlled chirality grown using designed catalysts},\ }\href {\doibase
  10.1038/nature21051} {\bibfield  {journal} {\bibinfo  {journal} {Nature}\
  }\textbf {\bibinfo {volume} {543}},\ \bibinfo {pages} {234} (\bibinfo {year}
  {2017})}\BibitemShut {NoStop}%
\bibitem [{\citenamefont {Yulaev}\ \emph {et~al.}(2016)\citenamefont {Yulaev},
  \citenamefont {Cheng}, \citenamefont {Walker}, \citenamefont {Vlassiouk},
  \citenamefont {Myers}, \citenamefont {Leite},\ and\ \citenamefont
  {Kolmakov}}]{Yulaev:2016}%
  \BibitemOpen
  \bibfield  {author} {\bibinfo {author} {\bibfnamefont {A.}~\bibnamefont
  {Yulaev}}, \bibinfo {author} {\bibfnamefont {G.}~\bibnamefont {Cheng}},
  \bibinfo {author} {\bibfnamefont {A.~R.~H.}\ \bibnamefont {Walker}}, \bibinfo
  {author} {\bibfnamefont {I.~V.}\ \bibnamefont {Vlassiouk}}, \bibinfo {author}
  {\bibfnamefont {A.}~\bibnamefont {Myers}}, \bibinfo {author} {\bibfnamefont
  {M.~S.}\ \bibnamefont {Leite}}, \ and\ \bibinfo {author} {\bibfnamefont
  {A.}~\bibnamefont {Kolmakov}},\ }\bibfield  {title} {\bibinfo {title} {Toward
  clean suspended {CVD} graphene},\ }\href {\doibase 10.1039/c6ra17360h}
  {\bibfield  {journal} {\bibinfo  {journal} {{RSC} Advances}\ }\textbf
  {\bibinfo {volume} {6}},\ \bibinfo {pages} {83954} (\bibinfo {year}
  {2016})}\BibitemShut {NoStop}%
\bibitem [{\citenamefont {Chen}\ \emph {et~al.}(2017)\citenamefont {Chen},
  \citenamefont {Stekovic}, \citenamefont {Li}, \citenamefont {Arkook},
  \citenamefont {Haddon},\ and\ \citenamefont {Bekyarova}}]{Chen:2017}%
  \BibitemOpen
  \bibfield  {author} {\bibinfo {author} {\bibfnamefont {M.}~\bibnamefont
  {Chen}}, \bibinfo {author} {\bibfnamefont {D.}~\bibnamefont {Stekovic}},
  \bibinfo {author} {\bibfnamefont {W.}~\bibnamefont {Li}}, \bibinfo {author}
  {\bibfnamefont {B.}~\bibnamefont {Arkook}}, \bibinfo {author} {\bibfnamefont
  {R.~C.}\ \bibnamefont {Haddon}}, \ and\ \bibinfo {author} {\bibfnamefont
  {E.}~\bibnamefont {Bekyarova}},\ }\bibfield  {title} {\bibinfo {title}
  {Sublimation-assisted graphene transfer technique based on small polyaromatic
  hydrocarbons},\ }\href {\doibase 10.1088/1361-6528/aa72d5} {\bibfield
  {journal} {\bibinfo  {journal} {Nanotechnology}\ }\textbf {\bibinfo {volume}
  {28}},\ \bibinfo {pages} {255701} (\bibinfo {year} {2017})}\BibitemShut
  {NoStop}%
\bibitem [{\citenamefont {Chen}\ \emph {et~al.}(2001)\citenamefont {Chen},
  \citenamefont {Zhang}, \citenamefont {Wang},\ and\ \citenamefont
  {Dai}}]{Chen:2001}%
  \BibitemOpen
  \bibfield  {author} {\bibinfo {author} {\bibfnamefont {R.~J.}\ \bibnamefont
  {Chen}}, \bibinfo {author} {\bibfnamefont {Y.}~\bibnamefont {Zhang}},
  \bibinfo {author} {\bibfnamefont {D.}~\bibnamefont {Wang}}, \ and\ \bibinfo
  {author} {\bibfnamefont {H.}~\bibnamefont {Dai}},\ }\bibfield  {title}
  {\bibinfo {title} {Noncovalent sidewall functionalization of single-walled
  carbon nanotubes for protein immobilization},\ }\href {\doibase
  10.1021/ja010172b} {\bibfield  {journal} {\bibinfo  {journal} {J. Am. Chem.
  Soc.}\ }\textbf {\bibinfo {volume} {123}},\ \bibinfo {pages} {3838} (\bibinfo
  {year} {2001})}\BibitemShut {NoStop}%
\bibitem [{\citenamefont {Lu}\ \emph {et~al.}(2006)\citenamefont {Lu},
  \citenamefont {Nagase}, \citenamefont {Zhang}, \citenamefont {Wang},
  \citenamefont {Ni}, \citenamefont {Maeda}, \citenamefont {Wakahara},
  \citenamefont {Nakahodo}, \citenamefont {Tsuchiya}, \citenamefont {Akasaka}
  \emph {et~al.}}]{Lu:2006}%
  \BibitemOpen
  \bibfield  {author} {\bibinfo {author} {\bibfnamefont {J.}~\bibnamefont
  {Lu}}, \bibinfo {author} {\bibfnamefont {S.}~\bibnamefont {Nagase}}, \bibinfo
  {author} {\bibfnamefont {X.}~\bibnamefont {Zhang}}, \bibinfo {author}
  {\bibfnamefont {D.}~\bibnamefont {Wang}}, \bibinfo {author} {\bibfnamefont
  {M.}~\bibnamefont {Ni}}, \bibinfo {author} {\bibfnamefont {Y.}~\bibnamefont
  {Maeda}}, \bibinfo {author} {\bibfnamefont {T.}~\bibnamefont {Wakahara}},
  \bibinfo {author} {\bibfnamefont {T.}~\bibnamefont {Nakahodo}}, \bibinfo
  {author} {\bibfnamefont {T.}~\bibnamefont {Tsuchiya}}, \bibinfo {author}
  {\bibfnamefont {T.}~\bibnamefont {Akasaka}},  \emph {et~al.},\ }\bibfield
  {title} {\bibinfo {title} {Selective interaction of large or charge-transfer
  aromatic molecules with metallic single-wall carbon nanotubes: critical role
  of the molecular size and orientation},\ }\href
  {https://pubs.acs.org/doi/abs/10.1021/ja058214%2B} {\bibfield  {journal}
  {\bibinfo  {journal} {J. Am. Chem. Soc.}\ }\textbf {\bibinfo {volume}
  {128}},\ \bibinfo {pages} {5114} (\bibinfo {year} {2006})}\BibitemShut
  {NoStop}%
\bibitem [{\citenamefont {Sundar}\ \emph {et~al.}(2004)\citenamefont {Sundar},
  \citenamefont {Zaumseil}, \citenamefont {Podzorov}, \citenamefont {Menard},
  \citenamefont {Willett}, \citenamefont {Someya}, \citenamefont {Gershenson},\
  and\ \citenamefont {Rogers}}]{Sundar:2004}%
  \BibitemOpen
  \bibfield  {author} {\bibinfo {author} {\bibfnamefont {V.~C.}\ \bibnamefont
  {Sundar}}, \bibinfo {author} {\bibfnamefont {J.}~\bibnamefont {Zaumseil}},
  \bibinfo {author} {\bibfnamefont {V.}~\bibnamefont {Podzorov}}, \bibinfo
  {author} {\bibfnamefont {E.}~\bibnamefont {Menard}}, \bibinfo {author}
  {\bibfnamefont {R.~L.}\ \bibnamefont {Willett}}, \bibinfo {author}
  {\bibfnamefont {T.}~\bibnamefont {Someya}}, \bibinfo {author} {\bibfnamefont
  {M.~E.}\ \bibnamefont {Gershenson}}, \ and\ \bibinfo {author} {\bibfnamefont
  {J.~A.}\ \bibnamefont {Rogers}},\ }\bibfield  {title} {\bibinfo {title}
  {Elastomeric transistor stamps: reversible probing of charge transport in
  organic crystals},\ }\href {\doibase 10.1126/science.1094196} {\bibfield
  {journal} {\bibinfo  {journal} {Science}\ }\textbf {\bibinfo {volume}
  {303}},\ \bibinfo {pages} {1644} (\bibinfo {year} {2004})}\BibitemShut
  {NoStop}%
\bibitem [{\citenamefont {Meitl}\ \emph {et~al.}(2005)\citenamefont {Meitl},
  \citenamefont {Zhu}, \citenamefont {Kumar}, \citenamefont {Lee},
  \citenamefont {Feng}, \citenamefont {Huang}, \citenamefont {Adesida},
  \citenamefont {Nuzzo},\ and\ \citenamefont {Rogers}}]{Meitl:2005}%
  \BibitemOpen
  \bibfield  {author} {\bibinfo {author} {\bibfnamefont {M.~A.}\ \bibnamefont
  {Meitl}}, \bibinfo {author} {\bibfnamefont {Z.-T.}\ \bibnamefont {Zhu}},
  \bibinfo {author} {\bibfnamefont {V.}~\bibnamefont {Kumar}}, \bibinfo
  {author} {\bibfnamefont {K.~J.}\ \bibnamefont {Lee}}, \bibinfo {author}
  {\bibfnamefont {X.}~\bibnamefont {Feng}}, \bibinfo {author} {\bibfnamefont
  {Y.~Y.}\ \bibnamefont {Huang}}, \bibinfo {author} {\bibfnamefont
  {I.}~\bibnamefont {Adesida}}, \bibinfo {author} {\bibfnamefont {R.~G.}\
  \bibnamefont {Nuzzo}}, \ and\ \bibinfo {author} {\bibfnamefont {J.~A.}\
  \bibnamefont {Rogers}},\ }\bibfield  {title} {\bibinfo {title} {Transfer
  printing by kinetic control of adhesion to an elastomeric stamp},\ }\href
  {\doibase 10.1038/nmat1532} {\bibfield  {journal} {\bibinfo  {journal} {Nat.
  Mater.}\ }\textbf {\bibinfo {volume} {5}},\ \bibinfo {pages} {33} (\bibinfo
  {year} {2005})}\BibitemShut {NoStop}%
\bibitem [{\citenamefont {Ishii}\ \emph {et~al.}(2015)\citenamefont {Ishii},
  \citenamefont {Yoshida},\ and\ \citenamefont {Kato}}]{Ishii:2015}%
  \BibitemOpen
  \bibfield  {author} {\bibinfo {author} {\bibfnamefont {A.}~\bibnamefont
  {Ishii}}, \bibinfo {author} {\bibfnamefont {M.}~\bibnamefont {Yoshida}}, \
  and\ \bibinfo {author} {\bibfnamefont {Y.~K.}\ \bibnamefont {Kato}},\
  }\bibfield  {title} {\bibinfo {title} {Exciton diffusion, end quenching, and
  exciton-exciton annihilation in individual air-suspended carbon nanotubes},\
  }\href {\doibase 10.1103/PhysRevB.91.125427} {\bibfield  {journal} {\bibinfo
  {journal} {Phys. Rev. B}\ }\textbf {\bibinfo {volume} {91}},\ \bibinfo
  {pages} {125427} (\bibinfo {year} {2015})}\BibitemShut {NoStop}%
\bibitem [{\citenamefont {Otsuka}\ \emph {et~al.}(2018)\citenamefont {Otsuka},
  \citenamefont {Yamamoto}, \citenamefont {Inoue}, \citenamefont {Koyano},
  \citenamefont {Ukai}, \citenamefont {Yoshikawa}, \citenamefont {Xiang},
  \citenamefont {Chiashi},\ and\ \citenamefont {Maruyama}}]{Otsuka:2018}%
  \BibitemOpen
  \bibfield  {author} {\bibinfo {author} {\bibfnamefont {K.}~\bibnamefont
  {Otsuka}}, \bibinfo {author} {\bibfnamefont {S.}~\bibnamefont {Yamamoto}},
  \bibinfo {author} {\bibfnamefont {T.}~\bibnamefont {Inoue}}, \bibinfo
  {author} {\bibfnamefont {B.}~\bibnamefont {Koyano}}, \bibinfo {author}
  {\bibfnamefont {H.}~\bibnamefont {Ukai}}, \bibinfo {author} {\bibfnamefont
  {R.}~\bibnamefont {Yoshikawa}}, \bibinfo {author} {\bibfnamefont
  {R.}~\bibnamefont {Xiang}}, \bibinfo {author} {\bibfnamefont
  {S.}~\bibnamefont {Chiashi}}, \ and\ \bibinfo {author} {\bibfnamefont
  {S.}~\bibnamefont {Maruyama}},\ }\bibfield  {title} {\bibinfo {title}
  {Digital isotope coding to trace the growth process of individual
  single-walled carbon nanotubes},\ }\href {\doibase 10.1021/acsnano.8b01630}
  {\bibfield  {journal} {\bibinfo  {journal} {{ACS} Nano}\ }\textbf {\bibinfo
  {volume} {12}},\ \bibinfo {pages} {3994} (\bibinfo {year}
  {2018})}\BibitemShut {NoStop}%
\bibitem [{\citenamefont {Legrand}\ \emph {et~al.}(2013)\citenamefont
  {Legrand}, \citenamefont {Roquelet}, \citenamefont {Lanty}, \citenamefont
  {Roussignol}, \citenamefont {Lafosse}, \citenamefont {Bouchoule},
  \citenamefont {Deleporte}, \citenamefont {Voisin},\ and\ \citenamefont
  {Lauret}}]{Legrand:2013}%
  \BibitemOpen
  \bibfield  {author} {\bibinfo {author} {\bibfnamefont {D.}~\bibnamefont
  {Legrand}}, \bibinfo {author} {\bibfnamefont {C.}~\bibnamefont {Roquelet}},
  \bibinfo {author} {\bibfnamefont {G.}~\bibnamefont {Lanty}}, \bibinfo
  {author} {\bibfnamefont {P.}~\bibnamefont {Roussignol}}, \bibinfo {author}
  {\bibfnamefont {X.}~\bibnamefont {Lafosse}}, \bibinfo {author} {\bibfnamefont
  {S.}~\bibnamefont {Bouchoule}}, \bibinfo {author} {\bibfnamefont
  {E.}~\bibnamefont {Deleporte}}, \bibinfo {author} {\bibfnamefont
  {C.}~\bibnamefont {Voisin}}, \ and\ \bibinfo {author} {\bibfnamefont {J.~S.}\
  \bibnamefont {Lauret}},\ }\bibfield  {title} {\bibinfo {title} {Monolithic
  microcavity with carbon nanotubes as active material},\ }\href {\doibase
  10.1063/1.4801984} {\bibfield  {journal} {\bibinfo  {journal} {Appl. Phys.
  Lett.}\ }\textbf {\bibinfo {volume} {102}},\ \bibinfo {pages} {153102}
  (\bibinfo {year} {2013})}\BibitemShut {NoStop}%
\bibitem [{\citenamefont {He}\ \emph {et~al.}(2017)\citenamefont {He},
  \citenamefont {Hartmann}, \citenamefont {Ma}, \citenamefont {Kim},
  \citenamefont {Ihly}, \citenamefont {Blackburn}, \citenamefont {Gao},
  \citenamefont {Kono}, \citenamefont {Yomogida}, \citenamefont {Hirano},
  \citenamefont {Tanaka}, \citenamefont {Kataura}, \citenamefont {Htoon},\ and\
  \citenamefont {Doorn}}]{He:2017}%
  \BibitemOpen
  \bibfield  {author} {\bibinfo {author} {\bibfnamefont {X.}~\bibnamefont
  {He}}, \bibinfo {author} {\bibfnamefont {N.~F.}\ \bibnamefont {Hartmann}},
  \bibinfo {author} {\bibfnamefont {X.}~\bibnamefont {Ma}}, \bibinfo {author}
  {\bibfnamefont {Y.}~\bibnamefont {Kim}}, \bibinfo {author} {\bibfnamefont
  {R.}~\bibnamefont {Ihly}}, \bibinfo {author} {\bibfnamefont {J.~L.}\
  \bibnamefont {Blackburn}}, \bibinfo {author} {\bibfnamefont {W.}~\bibnamefont
  {Gao}}, \bibinfo {author} {\bibfnamefont {J.}~\bibnamefont {Kono}}, \bibinfo
  {author} {\bibfnamefont {Y.}~\bibnamefont {Yomogida}}, \bibinfo {author}
  {\bibfnamefont {A.}~\bibnamefont {Hirano}}, \bibinfo {author} {\bibfnamefont
  {T.}~\bibnamefont {Tanaka}}, \bibinfo {author} {\bibfnamefont
  {H.}~\bibnamefont {Kataura}}, \bibinfo {author} {\bibfnamefont
  {H.}~\bibnamefont {Htoon}}, \ and\ \bibinfo {author} {\bibfnamefont {S.~K.}\
  \bibnamefont {Doorn}},\ }\bibfield  {title} {\bibinfo {title} {Tunable
  room-temperature single-photon emission at telecom wavelengths from $sp^3$
  defects in carbon nanotubes},\ }\href
  {http://dx.doi.org/10.1038/nphoton.2017.119} {\bibfield  {journal} {\bibinfo
  {journal} {Nat. Photon.}\ }\textbf {\bibinfo {volume} {11}},\ \bibinfo
  {pages} {577} (\bibinfo {year} {2017})}\BibitemShut {NoStop}%
\bibitem [{\citenamefont {Schweiger}\ \emph {et~al.}(2015)\citenamefont
  {Schweiger}, \citenamefont {Zakharko}, \citenamefont {Gannott}, \citenamefont
  {Grimm},\ and\ \citenamefont {Zaumseil}}]{Schweiger:2015}%
  \BibitemOpen
  \bibfield  {author} {\bibinfo {author} {\bibfnamefont {M.}~\bibnamefont
  {Schweiger}}, \bibinfo {author} {\bibfnamefont {Y.}~\bibnamefont {Zakharko}},
  \bibinfo {author} {\bibfnamefont {F.}~\bibnamefont {Gannott}}, \bibinfo
  {author} {\bibfnamefont {S.~B.}\ \bibnamefont {Grimm}}, \ and\ \bibinfo
  {author} {\bibfnamefont {J.}~\bibnamefont {Zaumseil}},\ }\bibfield  {title}
  {\bibinfo {title} {Photoluminescence enhancement of aligned arrays of
  single-walled carbon nanotubes by polymer transfer},\ }\href {\doibase
  10.1039/c5nr05163k} {\bibfield  {journal} {\bibinfo  {journal} {Nanoscale}\
  }\textbf {\bibinfo {volume} {7}},\ \bibinfo {pages} {16715} (\bibinfo {year}
  {2015})}\BibitemShut {NoStop}%
\bibitem [{\citenamefont {Cambr{\'{e}}}\ \emph {et~al.}(2012)\citenamefont
  {Cambr{\'{e}}}, \citenamefont {Santos}, \citenamefont {Wenseleers},
  \citenamefont {Nugraha}, \citenamefont {Saito}, \citenamefont {Cognet},\ and\
  \citenamefont {Lounis}}]{Cambre:2012}%
  \BibitemOpen
  \bibfield  {author} {\bibinfo {author} {\bibfnamefont {S.}~\bibnamefont
  {Cambr{\'{e}}}}, \bibinfo {author} {\bibfnamefont {S.~M.}\ \bibnamefont
  {Santos}}, \bibinfo {author} {\bibfnamefont {W.}~\bibnamefont {Wenseleers}},
  \bibinfo {author} {\bibfnamefont {A.~R.~T.}\ \bibnamefont {Nugraha}},
  \bibinfo {author} {\bibfnamefont {R.}~\bibnamefont {Saito}}, \bibinfo
  {author} {\bibfnamefont {L.}~\bibnamefont {Cognet}}, \ and\ \bibinfo {author}
  {\bibfnamefont {B.}~\bibnamefont {Lounis}},\ }\bibfield  {title} {\bibinfo
  {title} {Luminescence properties of individual empty and water-filled
  single-walled carbon nanotubes},\ }\href {\doibase 10.1021/nn300035y}
  {\bibfield  {journal} {\bibinfo  {journal} {{ACS} Nano}\ }\textbf {\bibinfo
  {volume} {6}},\ \bibinfo {pages} {2649} (\bibinfo {year} {2012})}\BibitemShut
  {NoStop}%
\bibitem [{\citenamefont {Fang}\ \emph {et~al.}(2020)\citenamefont {Fang},
  \citenamefont {Otsuka}, \citenamefont {Ishii}, \citenamefont {Taniguchi},
  \citenamefont {Watanabe}, \citenamefont {Nagashio},\ and\ \citenamefont
  {Kato}}]{Fang:2020}%
  \BibitemOpen
  \bibfield  {author} {\bibinfo {author} {\bibfnamefont {N.}~\bibnamefont
  {Fang}}, \bibinfo {author} {\bibfnamefont {K.}~\bibnamefont {Otsuka}},
  \bibinfo {author} {\bibfnamefont {A.}~\bibnamefont {Ishii}}, \bibinfo
  {author} {\bibfnamefont {T.}~\bibnamefont {Taniguchi}}, \bibinfo {author}
  {\bibfnamefont {K.}~\bibnamefont {Watanabe}}, \bibinfo {author}
  {\bibfnamefont {K.}~\bibnamefont {Nagashio}}, \ and\ \bibinfo {author}
  {\bibfnamefont {Y.~K.}\ \bibnamefont {Kato}},\ }\bibfield  {title} {\bibinfo
  {title} {Hexagonal boron nitride as an ideal substrate for carbon nanotube
  photonics},\ }\href {\doibase 10.1021/acsphotonics.0c00406} {\bibfield
  {journal} {\bibinfo  {journal} {{ACS} Photonics}\ }\textbf {\bibinfo {volume}
  {7}},\ \bibinfo {pages} {1773} (\bibinfo {year} {2020})}\BibitemShut
  {NoStop}%
\bibitem [{\citenamefont {Suzuki}\ and\ \citenamefont
  {Kobayashi}(2007)}]{Suzuki:2007}%
  \BibitemOpen
  \bibfield  {author} {\bibinfo {author} {\bibfnamefont {S.}~\bibnamefont
  {Suzuki}}\ and\ \bibinfo {author} {\bibfnamefont {Y.}~\bibnamefont
  {Kobayashi}},\ }\bibfield  {title} {\bibinfo {title} {Healing of low-energy
  irradiation-induced defects in single-walled carbon nanotubes at room
  temperature},\ }\href {\doibase 10.1021/jp067398r} {\bibfield  {journal}
  {\bibinfo  {journal} {J. Phys. Chem. C}\ }\textbf {\bibinfo {volume} {111}},\
  \bibinfo {pages} {4524} (\bibinfo {year} {2007})}\BibitemShut {NoStop}%
\bibitem [{\citenamefont {Miura}\ \emph {et~al.}(2014)\citenamefont {Miura},
  \citenamefont {Imamura}, \citenamefont {Ohta}, \citenamefont {Ishii},
  \citenamefont {Liu}, \citenamefont {Shimada}, \citenamefont {Iwamoto},
  \citenamefont {Arakawa},\ and\ \citenamefont {Kato}}]{Miura:2014}%
  \BibitemOpen
  \bibfield  {author} {\bibinfo {author} {\bibfnamefont {R.}~\bibnamefont
  {Miura}}, \bibinfo {author} {\bibfnamefont {S.}~\bibnamefont {Imamura}},
  \bibinfo {author} {\bibfnamefont {R.}~\bibnamefont {Ohta}}, \bibinfo {author}
  {\bibfnamefont {A.}~\bibnamefont {Ishii}}, \bibinfo {author} {\bibfnamefont
  {X.}~\bibnamefont {Liu}}, \bibinfo {author} {\bibfnamefont {T.}~\bibnamefont
  {Shimada}}, \bibinfo {author} {\bibfnamefont {S.}~\bibnamefont {Iwamoto}},
  \bibinfo {author} {\bibfnamefont {Y.}~\bibnamefont {Arakawa}}, \ and\
  \bibinfo {author} {\bibfnamefont {Y.~K.}\ \bibnamefont {Kato}},\ }\bibfield
  {title} {\bibinfo {title} {Ultralow mode-volume photonic crystal nanobeam
  cavities for high-efficiency coupling to individual carbon nanotube
  emitters},\ }\href {\doibase 10.1038/ncomms6580} {\bibfield  {journal}
  {\bibinfo  {journal} {Nat. Commun.}\ }\textbf {\bibinfo {volume} {5}},\
  \bibinfo {pages} {5580} (\bibinfo {year} {2014})}\BibitemShut {NoStop}%
\bibitem [{\citenamefont {Qian}\ \emph {et~al.}(2008)\citenamefont {Qian},
  \citenamefont {Georgi}, \citenamefont {Anderson}, \citenamefont {Green},
  \citenamefont {Hersam}, \citenamefont {Novotny},\ and\ \citenamefont
  {Hartschuh}}]{Qian:2008}%
  \BibitemOpen
  \bibfield  {author} {\bibinfo {author} {\bibfnamefont {H.}~\bibnamefont
  {Qian}}, \bibinfo {author} {\bibfnamefont {C.}~\bibnamefont {Georgi}},
  \bibinfo {author} {\bibfnamefont {N.}~\bibnamefont {Anderson}}, \bibinfo
  {author} {\bibfnamefont {A.~A.}\ \bibnamefont {Green}}, \bibinfo {author}
  {\bibfnamefont {M.~C.}\ \bibnamefont {Hersam}}, \bibinfo {author}
  {\bibfnamefont {L.}~\bibnamefont {Novotny}}, \ and\ \bibinfo {author}
  {\bibfnamefont {A.}~\bibnamefont {Hartschuh}},\ }\bibfield  {title} {\bibinfo
  {title} {Exciton energy transfer in pairs of single-walled carbon
  nanotubes},\ }\href {\doibase 10.1021/nl080048r} {\bibfield  {journal}
  {\bibinfo  {journal} {Nano Lett.}\ }\textbf {\bibinfo {volume} {8}},\
  \bibinfo {pages} {1363} (\bibinfo {year} {2008})}\BibitemShut {NoStop}%
\bibitem [{\citenamefont {Ye}\ \emph {et~al.}(2018)\citenamefont {Ye},
  \citenamefont {Liu}, \citenamefont {Han}, \citenamefont {Ge}, \citenamefont
  {Cui}, \citenamefont {Zhang}, \citenamefont {Zheng}, \citenamefont {Liu},
  \citenamefont {Liu}, \citenamefont {Liu},\ and\ \citenamefont
  {Tao}}]{Ye:2018}%
  \BibitemOpen
  \bibfield  {author} {\bibinfo {author} {\bibfnamefont {X.}~\bibnamefont
  {Ye}}, \bibinfo {author} {\bibfnamefont {Y.}~\bibnamefont {Liu}}, \bibinfo
  {author} {\bibfnamefont {Q.}~\bibnamefont {Han}}, \bibinfo {author}
  {\bibfnamefont {C.}~\bibnamefont {Ge}}, \bibinfo {author} {\bibfnamefont
  {S.}~\bibnamefont {Cui}}, \bibinfo {author} {\bibfnamefont {L.}~\bibnamefont
  {Zhang}}, \bibinfo {author} {\bibfnamefont {X.}~\bibnamefont {Zheng}},
  \bibinfo {author} {\bibfnamefont {G.}~\bibnamefont {Liu}}, \bibinfo {author}
  {\bibfnamefont {J.}~\bibnamefont {Liu}}, \bibinfo {author} {\bibfnamefont
  {D.}~\bibnamefont {Liu}}, \ and\ \bibinfo {author} {\bibfnamefont
  {X.}~\bibnamefont {Tao}},\ }\bibfield  {title} {\bibinfo {title}
  {Microspacing in-air sublimation growth of organic crystals},\ }\href
  {\doibase 10.1021/acs.chemmater.7b04170} {\bibfield  {journal} {\bibinfo
  {journal} {Chem. Mater.}\ }\textbf {\bibinfo {volume} {30}},\ \bibinfo
  {pages} {412} (\bibinfo {year} {2018})}\BibitemShut {NoStop}%
\bibitem [{\citenamefont {Ohno}\ \emph {et~al.}(2006)\citenamefont {Ohno},
  \citenamefont {Iwasaki}, \citenamefont {Murakami}, \citenamefont {Kishimoto},
  \citenamefont {Maruyama},\ and\ \citenamefont {Mizutani}}]{Ohno:2006prb}%
  \BibitemOpen
  \bibfield  {author} {\bibinfo {author} {\bibfnamefont {Y.}~\bibnamefont
  {Ohno}}, \bibinfo {author} {\bibfnamefont {S.}~\bibnamefont {Iwasaki}},
  \bibinfo {author} {\bibfnamefont {Y.}~\bibnamefont {Murakami}}, \bibinfo
  {author} {\bibfnamefont {S.}~\bibnamefont {Kishimoto}}, \bibinfo {author}
  {\bibfnamefont {S.}~\bibnamefont {Maruyama}}, \ and\ \bibinfo {author}
  {\bibfnamefont {T.}~\bibnamefont {Mizutani}},\ }\bibfield  {title} {\bibinfo
  {title} {Chirality-dependent environmental effects in photoluminescence of
  single-walled carbon nanotubes},\ }\href {\doibase
  10.1103/PhysRevB.73.235427} {\bibfield  {journal} {\bibinfo  {journal} {Phys.
  Rev. B}\ }\textbf {\bibinfo {volume} {73}},\ \bibinfo {pages} {235427}
  (\bibinfo {year} {2006})}\BibitemShut {NoStop}%
\end{thebibliography}
\end{document}